\documentclass[12pt]{article}
\usepackage{amsmath,amssymb}

\renewcommand{\theequation}{\arabic{equation}}
\newcommand{\be}{\begin{equation}}
\newcommand{\ee}{\end{equation}}
\newcommand{\bea}{\begin{array}}
\newcommand{\ea}{\end{array}}
\newcommand{\beqa}{\begin{eqnarray}}
\newcommand{\eeqa}{\end{eqnarray}}
\newcommand{\bean}{\begin{eqnarray*}}
\newcommand{\eean}{\end{eqnarray*}}

\newcommand{\nn}{\nonumber}

\def\up#1{\leavevmode \raise.16ex\hbox{#1}}

\def\sqr#1#2{{\vcenter{\vbox{\hrule height.#2pt
        \hbox{\vrule width.#2pt height#1pt \kern#1pt
          \vrule width.#2pt}
        \hrule height.#2pt}}}}

\def\CP2{{\mathbb C}P^2}

\renewcommand{\theequation}{\thesection.\arabic{equation}}
\newcommand{\gapproxeq}{\lower .7ex\hbox{$\;\stackrel{\textstyle
>}{\sim}\;$}}
\newcommand{\lapproxeq}{\lower .7ex\hbox{$\;\stackrel{\textstyle
<}{\sim}\;$}}

\newcounter{appendice}
\newcommand{\appendice}
{
\setcounter{equation}{0}
\renewcommand{\theequation}{\Alph{appendice}.\arabic{equation}}
\addtocounter{appendice}{1}
{\bf Appendix \Alph{appendice}} 
}

\def\thebibliography#1{{\bf REFERENCES\markboth
 {REFERENCES}{REFERENCES}}\list
 {[\arabic{enumi}]}{\settowidth\labelwidth{[#1]}\leftmargin\labelwidth
 \advance\leftmargin\labelsep
 \usecounter{enumi}}
 \def\newblock{\hskip .11em plus .33em minus -.07em}
 \sloppy
 \sfcode`\.=1000\relax}

\begin{document}

\begin{titlepage}

\title{{\small\hfill SU-4240-730}\\ Fuzzy $\CP2$}

\author{G.Alexanian$^*$,
A.P.Balachandran$^*$,
G.Immirzi$^\dagger$
and B.Ydri$^*$\\
~\\
{\small\it $^*$ Physics Department, 
Syracuse University}\\
{\small\it Syracuse NY 13244, USA}\\
~\\
{\small\it $^\dagger$ Universita' di Perugia {\small\rm and} INFN}\\
{\small\it Perugia, Italy}\\
}

\maketitle

\begin{abstract}

Regularization of quantum field theories (QFT's) can be achieved
by quantizing the underlying manifold (spacetime or spatial slice)
thereby replacing it by a non-commutative matrix model or a 
``fuzzy manifold''.
Such discretization by quantization is remarkably successful in
preserving symmetries and topological features, and
altogether overcoming the fermion-doubling problem.
In this paper, we report on our work on the ``fuzzification''
of the four-dimensional $\CP2$ and its QFT's.
$\CP2$ is not spin, but spin${}_c$. Its Dirac operator has
many unique features. They are explained and their fuzzy versions
are described.
\end{abstract}
\end{titlepage}

\section{Introduction}

We can find few fundamental physical models amenable to
exact treatment. Approximation methods like perturbation theory
are necessary and are part of our physics culture.

Among the important approximation methods for quantum field theories
(QFT's) are strong coupling methods based on lattice discretisation
of underlying spacetime or perhaps its time-slice.
They are among the rare effective approaches for the study
of confinement in QCD and for non-perturbative regularization
of QFT's.
They enjoyed much popularity in their early days and have retained
their good reputation for addressing certain fundamental problems.

One feature of naive lattice discretisations however can be criticised.
They do not retain the symmetries of the exact theory
except in some rough sense. A related feature is that topology and
differential geometry of the underlying manifolds are treated only 
indirectly, by limiting the couplings to ``nearest neighbours''.
Thus lattice points are generally manipulated  like a trivial 
topological set, with a point being both open
and closed. The upshot is that these models have  no rigorous
representation of topological defects and lumps like vortices,
solitons and monopoles. The complexities in the ingenious solutions
for the discrete QCD $\theta$-term \cite{ref1} illustrate such
limitations.
There do exist radical attempts to overcome these limitations
using partially ordered sets \cite{ref2}, but their
potentials are yet to be adequately studied.

A new approach to discretisation, inspired by non-commutative
geometry (NCG), is being developed since a few 
years [4-15].
The key remark here is that when the underlying spacetime
or spatial cut can be treated as a phase space and quantized,
with a parameter $\hat h$ assuming the role of $\hbar$,
the emergent quantum space is fuzzy, and the number of 
independent states per (``classical'') unit volume
becomes finite. We have known this result
after Planck and Bose introduced such an unltraviolet
cut-off and quantum physics later justified it.
``Fuzzified'' manifold is ultraviolet finite, and if
the parent manifold is compact too, supports only finitely 
many independent states. The continuum limit is the semiclassical 
$\hat h\rightarrow 0$ limit.
This unconventional discretization of classical topology
is not at all equivalent to the naive one, and we shall
see that it does significantly overcome the previous
criticisms. 

There are other reasons also to pay attention to
fuzzy spaces, be they spacetimes or spatial cuts.
There is much interest among string theorists in matrix
models and in describing D-branes using matrices. Fuzzy spaces
lead to matrix models too and their ability to reflect topology
better than elsewhere should therefore evoke our curiosity.
They let us devise new sorts of discrete models and are 
interesting from that perspective.
In addition, it has now been discovered that when open strings 
end on D-branes which are symplectic manifolds, then the branes
can \cite{ref5} become fuzzy, in this way one comes across fuzzy
tori, ${\mathbb C}P^N$ and many such spaces in string physics.

The central idea behind fuzzy spaces is discretisation by 
quantization.
It does not always work. An obvious limitation is that
the parent manifold  has to be even dimensional. 
(See  however ref. \cite{ref6} for fuzzyfying $RP^3/{\mathbb Z}_2$ and other 
non-symplectic manifolds, even or odd).
If it is not, it has no chance of being a phase space. But
that is not all. Successful use of fuzzy spaces
for QFT's requires good fuzzy versions of
the Laplacian, Dirac equation, chirality operator
and so forth, and their incorporation can make the entire
enterprise complicated. The torus $T^2$ is compact,
admits a symplectic structure and on quantization
becomes fuzzy, or a non-commutative  torus. It supports a
finite number of states if the symplectic form satisfies
the Dirac quantization condition. But it is impossible to 
introduce suitable derivations without escalating
the formalism to infinite dimensions \cite{ref7}.

But we do find a family of classical manifolds elegantly
escaping these limitations. They are the co-adjoint orbits
of Lie groups. For semi-simple Lie groups, they are the same 
as adjoint orbits. It is a theorem that these orbits 
are symplectic \cite{ref8}. They can often be 
quantized when the 
symplectic forms satisfy the Dirac quantization condition.
The resultant fuzzy spaces are described by linear operators
on irreducible representations (IRR's) of the group.
For compact orbits, the latter are finite-dimensional.
In addition, the elements of the Lie algebra define natural
derivations, and that helps to find Laplacian and the Dirac 
operator. We can even define chirality with no fermion 
doubling and represent monopoles and instantons (See [4-9]
and the first 3 papers in \cite{ref9}).
These orbits therefore are 
altogether well-adapted for QFT's.

Let us give examples of these orbits:
\begin{itemize}
\item{$S^2$}:
This is the orbit of $SU(2)$ through the Pauli matrix 
$\sigma_3$ or any of its multiples $\lambda\,\sigma_3$
($\lambda\neq 0$). It is the set  
$\{\lambda\,g\,\sigma_3\,g^{-1}\,:\, g\in SU(2)\}$.
The symplectic form is $j\,d\,{\rm cos}\,\theta\wedge d\phi$
with $\theta,\phi$ being the usual $S^2$ coordinates,
\cite{ref10}. Quantization gives the spin $j$ $SU(2)$
representations.
\item{$\CP2$}:
$\CP2$ is of particular interest being of dimension 4.
It is the orbit of $SU(3)$ through the hypercharge
$Y=1/3\,\,{\rm diag}(1,1,-2)$ (or its multiples):
\begin{equation}
\CP2:\{g\,Y\,g^{-1}\,:\, g\in SU(3)\}.
\end{equation}
The associated representations are symmetric products
of $3$'s or ${\bar 3}$'s. (See section 2).
\item{$SU(3)/[U(1)\times U(1)]$}:
This 6-dimensional manifold is the orbit of $SU(3)$
through $\lambda_3={\rm diag}(1,-1,0)$ and its multiples.
These orbits give all the IRR's containing a zero hypercharge state.
\end{itemize}
In the literature, there are several studies of the
fuzzy physics of ${\mathbb C}P^1=S^2$ [3-15] while there is
also a rigorous and beautiful treatment of $CP^2$ by Grosse
and Strohmaier \cite{ref11}. The present work develops an 
alternative formulation for $\CP2$. It is close to earlier 
treatments of $S^2$ \cite{ref12,ref9} and seems to generalize 
to other quantizable orbits. It is eventually 
equivalent to that of \cite{ref11} as we show, so that
the first study of  $\CP2$ is of that reference.

Throughout this paper, we treat $\CP2$ as 
Euclidean spacetime even though the possibility of treating 
it as spacial slice is also available. 

Section 2 explains the basic properties of $\CP2$.
We quantize it in Section 3 to produce the fuzzy
$\CP2$ (Some technical details necsessary
to quantization are provided in the Appendix A). 
Functional integral quantization of tensorial fields 
can also be done as we show in Section 4
(although topological considerations would prefer a 
more elaborate approach;
see especially the first and last papers of ref.
\cite{ref9}).
In non-commutative geometry (NCG), a central role is assumed 
by the (massless) Dirac operator. Section 5 reviews it
for $S^2={\mathbb C}P^1$ while Section 6 studies our approach to it
in detail for $\CP2$.
Analysis shows its equivalence to the Dirac-K{\"a}hler operator
\cite{ref11}.  $\CP2$ is not a spin, but a spin${}_c$ manifold,
and that has exotic consequences for the $SU(3)$
spectrum: left and right chiral modes transform differently
under $SU(3)$. Section 7 studies the fuzzy analogue
of the Dirac operator.
This work is greatly facilitated
by coherent states and star ($\star$) products. The necessary
material, contained in \cite{ref6,ref13}, is reviewed
and used to discretise the continuum  material here for both
$S^2={\mathbb C}P^1$ and $\CP2$.
Incidentally the  $\star$ product is particularly useful
for formulating fuzzy analogues of important continuum
quantities like correlation functions.

$\CP2$ is a surface in ${\mathbb R}^8$ described by an
algebraic equation. Appendix A establishes the fuzzy version
of this equation and in addition useful identities among
$SU(3)$ generators.

Appendix B is pedagogical and explains
why $\CP2$ is not spin and why the $SU(3)$ spectrum
of the spin${}_c$ Dirac operator has exotic features.

\section{On $\CP2$}
\setcounter{equation}{0}

$\CP2$ is a K{\"a}hler manifold describable 
in different ways. Thus, as mentioned before
it is the orbit of $SU(3)$ through the hypercharge
operator $Y$ or its multiples (the group $SU(3)$ has eight generators 
$t_i$ which satisfy $[t_i,t_j]=if_{ijk}t_k$; 
the hypercharge is $Y=\frac{2}{\sqrt 3}t_8$; in the {\bf 3} 
representation the generators are $\frac{1}{2}\lambda_i$, where the 
$\lambda_i$ are the eight Gell-Mann matrices).
As the stability group of $Y$ is $U(2)$:
\begin{equation}
U(2)=\left\{\left(
\begin{array}{cc}
u & 0\\
0 & {det\,u^{-{1}}}\\
\end{array}
\right) \in SU(3)\right\},
\end{equation}
we have that 
\begin{equation}
\CP2=SU(3)/U(2).
\label{s21}
\end{equation}

As its name reveals, it is also a projective complex
space or the space of $C^1$ subspaces in $C^3$.
If $\xi\in C^3-\{0\}$, a point of $\CP2$ is the equivalence class
$\langle\xi\rangle=\langle\lambda\xi\rangle$
for all $\lambda\in C^1-\{0\}$.
Choosing $\lambda=\left(\sum|\xi_i|^2\right)^{-\frac{1}{2}}$,
we see that 
$\CP2=\{\langle\xi\rangle=\langle\xi e^{i\theta}\rangle$:
$\left(\sum|\xi_i|^2\right)=1\}$.
Hence 
\begin{equation}
\CP2=S^5/U(1).
\end{equation}
In (\ref{s21}), we can first quotient $SU(3)$ by $SU(2)$.
That is just the above $S^5$. That is because $SU(3)$ acts
on $C^3$ and transitively on its sphere 
$S^5 = \{\zeta \in C^3 :\sum |\zeta_i|^2=1\}$.
At $(1,0,0)\in S^5$, the stability group is $SU(2)$ 
showing the result. In this way we see that
\begin{equation}
\CP2=[SU(3)/SU(2)]/U(1)=S^5/U(1).
\label{s22}
\end{equation}

The eight Gell-Mann matrices form a basis for the real vector 
space of traceless hermitian matrices
$\{\sum\xi_i\lambda_i, \xi=(\xi_1,...,\xi_8)\in R^8\}$.
So $\CP2$ is a submanifold of $R^8$.
There is a beautiful algebraic equation for this 
submanifold. It is this:
Let $d_{ijk}$ be the totally symmetric $SU(3)$-invariant tensor
defined by 
 \begin{equation}
 \lambda_i\lambda_j=\frac{2}{3}\,\delta_{ij}+(\,d_{ijk}+if_{ijk})\,\lambda_k
 \label{s23}
 \end{equation}
Then
\begin{equation}
\xi\in \CP2\ \ \ \Longleftrightarrow\ \ \ 
d_{ijk}\xi_i\xi_j=\,{\rm constant}\times\,\xi_k.
\label{s24}
\end{equation}
A pleasant manner to demonstrate this result is as follows.
The symmetric $SU(3)$ invariant product 
$\chi,\eta \rightarrow \chi\vee\eta, (\chi\vee\eta)_i :=
d_{ijk} \chi_j\,\eta_k$ can be rewritten in terms of traceless 
hermitian matrices $M,N$ as
\begin{equation}
M\vee N = \frac{1}{2}\{M,N\}-\frac{1}{6}{\rm Tr}\left(\{M,N\}\right),
\label{s25}
\end{equation}
where
\begin{equation}
M=\sum\chi_j\lambda_j,\ \ \ N=\sum\eta_j\lambda_j.
\end{equation}
For this product, $M=\delta \lambda_8$ fulfills
\begin{equation}
M\vee M=-\frac{\delta}{\sqrt{3}}M
\label{s26}
\end{equation}
This relation is valid for points on the entire orbit
through $\delta\lambda_8$ by $SU(3)$ invariance of the $\vee$
product.

Conversely, relation (\ref{s26}) implies that $M$ is in the orbit of 
$\lambda_8$. For we can diagonalize $M$ by an $SU(3)$ transformation
$g$ while keeping (\ref{s26}). After scaling 
the diagonal $M_D=g\,M\,g^{-1}$
to $\Delta$ to reduce $-\frac{\delta}{\sqrt{3}}$ to 1,
we have 
$\Delta \vee \Delta=\Delta$,
$\Delta= {\rm diag}(a,b,-a-b)$.
Comparing the difference  of the first two rows on both sides, 
we get $a-b=(a+b)(a-b)$.
If $a=b$, then $\Delta=3\,a\,Y$. If $a\neq b$, then $a+b=1$. 
Comparing the first row, we get  
$a^2-a-2=0$, or $a=2$ or $-1$. So 
$\Delta = {\rm diag}(2,-1,-1)$ or  $\Delta = {\rm diag}(-1,2,-1)$.
Both become proportional to $Y$ after Weyl reflections,
establishing the result.

\section{Quantizing $\CP2$}
\setcounter{equation}{0}

A particular approach to quantizing coadjoint orbits was developed 
many years ago in \cite{ref10}.
According to that method we obtain  fuzzy $\CP2$ quantizing
the Lagrangian:
\begin{equation}
L=i\,{\bar N}{\rm Tr} Y\,g(t)^{-1}\,{\dot g(t)},\ \ \ \ \
g(t)\in SU(3), \ \ {\bar N}={\rm constant},\ \ Y=\frac{\lambda_8}{\sqrt 3}
\label{s31}
\end{equation}
A point $\xi(t) \in \CP2$ is related to $g(t)$ by
$\xi(t)_i\,\lambda_i\,=\,g(t)\,Y\,g^{-1}(t)$, while 
on $\CP2$ the symplectic form is 
$i\,{\bar N}\,d\,{\rm Tr}\,Y\,g^{-1}\,dg\,=
-i\,{\bar N}{\rm Tr}Y\left[g^{-1}dg\wedge g^{-1}dg\right]$.
Writing
$g=e^{i\lambda_i\theta^i/2}$, for a Hamiltonian description we may take as 
phase space (local)
coordinates the $\theta^i$ and their conjugates $\pi_i=
\frac{\partial L}{\partial\dot\theta^i}$, but the Lagrangian being of first 
order the latter are all constraints. To simplify 
the constraints we define $E_{ij}$ 
by $g^{-1}dg=\frac{\lambda_j}{2i}E_{ji}d\theta^i$, and  use
the variables $\Lambda_{iR}=-\pi_j(E^{-1})_{ji}$, which have Poisson brackets:
\begin{equation}
\{ \Lambda_{iR},g \} =g\frac{\lambda_i}{2i}\ \ ;\ \ \
\{ \Lambda_{iR},\Lambda_{jR} \} =f_{ijk}\Lambda_{kR}
\end{equation} 
and are therefore the generators of $SU(3)$ transformations on $g(t)$
acting on the right. In terms of these variables the constraints 
$\pi_i-i\bar N{\rm Tr}(Yg^{-1}\frac{\partial g}{\partial \theta^i})
\approx 0$ become: 
\begin{equation}
\Lambda_{iR}+\frac{\bar N}{\sqrt 3}\delta_{i,8}\approx 0
\end{equation}
They are second class for $i=4,..,7$, first class for $i=1,2,3,8$, 
corresponding to the fact that if $g(t)\rightarrow g(t)e^{i\lambda_i
\theta(t)/2},\ i=1,2,3,8$, then 
$L\rightarrow L-\frac{\bar N}{\sqrt 3}\dot\theta\delta_{i,8}$.
Thus for the generator $Y_R$ for right hypercharge we have 
$Y_R\approx -\frac{2}{3}{\bar N}$, and
the right ''isospin'' generators $I_{\alpha R}\;,\alpha=1,2, 3$
vanish, $I_{\alpha R}\approx 0$.
We can make a first class set
(classically equivalent to all the constraints) by adding to these
constraints complex combinations of the second class constraints:
$Y_R\approx-\frac{2}{3}{\bar N}, I_{\alpha R}\approx 0$
and for ${\bar N}\ge 0$, 
$\Lambda_{4R}-i\Lambda_{5R}\approx\Lambda_{6R}-i\Lambda_{7R}\approx 0$
and for  ${\bar N}\le 0$,
$ \Lambda_{4R}+i\Lambda_{5R}\approx\Lambda_{6R}+i\Lambda_{7R}\approx 0$.

These constraints can be realized on functions on $SU(3)$.
As isospin singlets have hypercharge in integral
multiples of $\frac{2}{3}$, we find that ${\bar N}\in {\mathbb Z}$.
With ${\bar N}$ fixed accordingly, the constraints together
mean that for right action, we have highest weight isospin
singlet states of  hypercharge  $-\frac{2}{3}{\bar N}$.

An IRR of $SU(3)$ is labeled by $(n_1,n_2), n_i\in {\mathbb N}$.
It comes from the symmetric product of $n_1$ {\underline 3}'s
and $n_2$ ${\underline 3}^*$'s: A tensor 
$T^{i_1...i_{n_1}}_{j_1...j_{n_2}}$ for  $(n_1,n_2)$ has $n_1$
upper indices, $n_2$ lower indices and is traceless, 
$T^{i_1i_2...i_{n_1}}_{i_1j_2...j_{n_2}}=0$.
Within an IRR, the orthonormal basis can be written as
$|(n_1,n_2), I^2,I_3,Y\rangle$
where $I^2,I_3$ and $Y$ are square of isospin,
its third component and hypercharge.

Let $g\rightarrow U^{(n_1,n_2)}(g)$ define the representation $(n_1,n_2)$
of $SU(3)$. Then the functions given by
$\langle(n_1,n_2),I^2,I_3,Y|U^{(n_1,n_2)}(g)|(n_1,n_2),0,0,
-\frac{2}{3}{\bar N}\rangle$
fulfill the constraints.
By the Peter-Weyl theorem, their linear span
\begin{equation}
\sum\xi^{(n_1,n_2)}_{I^2,I_3,Y}
\langle(n_1,n_2),I^2,I_3,Y|U^{(n_1,n_2)}(g)|(n_1,n_2),0,0,
-{\textstyle\frac{2}{3}}{\bar N}\rangle
\nonumber
\end{equation} 
gives all the functions of interest.

If ${\bar N}=N\ge 0$, that requires that $(n_1,n_2)=(N,0)$.
These are just the symmetric products of $N$ ${\underline 3}$'s.
If  ${\bar N}=-N\le 0$, $(n_1,n_2)=(0,N)$ or we get the symmetric
product of $N$ ${\underline 3}^*$'s.
The representations that we get by quantizing the
Lagrangian (\ref{s31}) are thus $(N,0)$ or $(0,N)$.

For $\CP2$, there are coordinate functions ${\hat\xi}_i$
where ${\hat\xi}_i(\xi)=\xi_i$. $\sum{\hat\xi}_i\,{\hat\xi}_i$
is a constant function which we can take to be ${\mathbb I}$,
the function with value one.  On quantization, ${\hat\xi}_i$ become 
the operators constant$\times\Lambda^L_i$ 
which we also denote as  ${\hat\xi}_i$.
Since $\sum\,\Lambda^L_i\,\Lambda^L_i=C_2{\mathbb I}$, and
$C_2=\frac{1}{3}N^2+N$ in $(N,0)$ or $(0,N)$ (see Appendix A),
their exact form is 
\begin{equation}
{\hat\xi}_i=\frac{\Lambda^L_i}{\sqrt{\frac{1}{3}N^2+N}},\ \ \ \
\sum{\hat\xi}_i\,{\hat\xi}_i={\mathbb I}.
\label{s32}
\end{equation}
So
\begin{equation}
\left[{\hat\xi}_i,{\hat\xi}_j\right]=\frac{i}{\sqrt{\frac{1}{3}N^2+N}}
\,f_{ijk}\,{\hat\xi}_k
\label{s33}
\end{equation}
and they commute in the large $N$ limit.

It is a remarkable fact that  ${\hat\xi}_i$ fulfill
(\ref{s24}) 
for {\it any} $N$ if $\hat{\xi}_i$'s belong to $(N,0)$ or $(0,N)$.  
A proof that uses the creation-annihilation
operator techniques of Grosse, Klim{\v c}ik and 
Pre\v{s}najder [6-9] is given in Appendix A.
The result is a ``fuzzy'' analog of the defining relation
(\ref{s24}):
\begin{equation}
d_{ijk}\hat{\xi}_i\hat{\xi}_j=\frac{\frac{N}{3}+\frac{1}{2}}{\sqrt{\frac{1}{3}N^2+N}}
\times\,\hat{\xi}_k.
\label{fuzzys24} 
\end{equation}

The algebra $A$ generated by ${\hat \xi}_i$ is what
substitutes for the algebra of functions ${\cal A}=C^\infty(\CP2)$.
By Burnside's theorem \cite{ref18}, it is the full matrix
algebra in the IRR.
Fuzzy $\CP2$ is just the algebra $A$.

The following point, emphasised by \cite{ref11}
is noteworthy. If ${f}\in {\cal A}$, it has the partial-wave
expansion
\begin{eqnarray}
&&{f(\xi)}=\sum{f}^n_{I^2,I_3,Y}
\langle(n_1,n_2),I^2,I_3,Y|U^{(n_1,n_2)}(g)|(n_1,n_2),0,0,0\rangle
\label{s34},\\
&&\xi_\alpha\,\lambda_\alpha\,:=\,g\,\lambda_8\,g^{-1}.
\nn
\end{eqnarray}
The ket $|(n_1,n_2),0,0,0\rangle $ exists only if $n_1=n_2$
so that the sum in (\ref{s34}) can be restricted to  $n_1=n_2$.
If $F\in A$, then $F$ too has an expansion like (\ref{s34})
where the series is cut-off at $n=N$. That is because of the
following. The $SU(3)$ Lie algebra $su(3)$ has two actions on $F$:
$F\rightarrow L^L_\alpha\,F=\Lambda_\alpha\,F$
and $F\rightarrow -L^R_\alpha\,F\,=-F\,\Lambda_\alpha$.
The derivation 
$F\rightarrow ad\, L_\alpha\,F=L^L_\alpha\,F
-L^R_\alpha\,F=\left[\Lambda_\alpha,F\right]$
is the action which annihilates ${\mathbb I}$
and corresponds to the $su(3)$ action on $\CP2$.
As $F$ transforms as $(N,0)$ (for ${\bar N}\ge 0$ say)
for $\Lambda^L_\alpha$ and as $(0,N)$ for $-\Lambda^R_\alpha$,
$A$ decomposes into direct sum of IRR's:
$(N,0)\otimes (0,N)=\oplus^N_{n=0}(n,n)$.
If $\langle(n,n),I^2,I_3,Y\rangle$ furnishes a basis for $(n,n)$,
then $F=\sum^N_0 F^n_{I^2,I_3,Y}|(n,n),I^2,I_3,Y\rangle$.
Identifying this basis with the one in (\ref{s34}) for $n\leq N$,
we see that $F$ transforms like a function on $\CP2$ with a 
terminating partial wave expansion.

A more precise statement is as follows \cite{ref11}.
We can put a scalar product on $\cal A$ using the Haar measure
on $SU(3)$ and complete $\cal A$ into a Hilbert space ${\cal H}$. 
On ${\cal H}$, 
elements $\cal F$ of 
$\cal A$ act as linear operators by point-wise multiplication. 
Let ${\cal H}_{(N,0)}$ be the subspace of ${\cal H}$ carrying the IRR $(N,0)$
and $P_{(N,0)}: {\cal H}\rightarrow {\cal H}_{(N,0)}$ the corresponding 
projector.
Then we have a map ${\cal A}\rightarrow P_{(N,0)}\,{\cal A}\, P_{(N,0)}$;
${\cal F}\rightarrow P_{(N,0)}\,{\cal F}\, P_{(N,0)}$ which is onto $A$.
Thus elements of $A$ approximate functions in a good sense.

\section{Fuzzy Scalar Fields}
\setcounter{equation}{0}


Here we briefly indicate a certain fuzzy version of the free
scalar field action. It is very natural and a generalization of
fuzzy ${\mathbb C}P^1$ action proposed earlier [5-11]. 
Still certain less obvious
actions based on cyclic cohomology have been proposed
\cite{ref9,ref6}, they have
distinct topological advantages and correct continuum limits as
well.

The operators ${ad\, L}_i=L_i^L-L_i^R$ correspond to the $SU(3)$
generators for functions on $\CP2$. A Laplacian for fuzzy $\CP2$
is thus ${ad\, L}_i^2$. A scalar field ${\phi}$ is a polynomial
in the fuzzy coordinate functions $\hat{{\xi}}_i$, so ${\phi}$
is just a matrix in $A$. The Euclidean action for ${\phi}$ is
\begin{eqnarray}
S(\phi)&=&{\rm ~constant}\times Tr({\phi}^{+}{ad\, L_i}^2{\phi}),\nonumber\\
{ad\, L}_i{\phi}&=&[L_i,\phi].
\end{eqnarray}

Let ${\lambda}_K$ be the eigenvalue of the continuum operator for
the IRR $(K,K)$. Ref. \cite{ref11} gives
\begin{equation}
\lambda_K=2K(K+1).
\end{equation}
If $N$ is the maximum $K$ for
the fuzzy space, then ${ad\, L}_i^2$ has the spectrum
$\{{\lambda}_0,{\lambda}_1,..{\lambda}_N\}$, it is just the
cut-off spectrum of the continuum Laplacian.

\section{The Dirac Operator on ${\bf S^2{\simeq}{\mathbb C}P^1}$}
\setcounter{equation}{0}
 
\noindent
This section is a warm up for what follows on $\CP2$ next. It
contains a partial-wave analysis for the eigenstates of the $S^2$
Dirac operator ${\cal D}$ which can be generalised to $\CP2$.

Let
\begin{equation}
S^2=\{x{\in}R^3:\sum x_\alpha^2=1\},
\label{sphere}
\end{equation}
and $\hat{x}$ be the coordinate functions:
$\hat{x}_{\alpha}(x)=x_{\alpha}$. 

Then the Dirac operator is
\begin{eqnarray}
{\cal D}&=&{\sigma}_{\alpha}{\cal
P}_{{\alpha}{\beta}}\,J_{\beta},\nonumber\\
{\cal
P}_{{\alpha}{\beta}}&=&{\delta}_{{\alpha}{\beta}}-\hat{x}_{\alpha}\hat{x}_{\beta},
J_{\beta}={\cal
L}_{\beta}+\frac{{\sigma}_{\beta}}{2},\nonumber\\ 
{\cal L}_\alpha&=&-i(\hat{x}{\wedge}\vec{\nabla})_{\alpha}.
\end{eqnarray}
${\cal P}$ projects the Pauli matrices ${\sigma}_\alpha$ to their
tangent space components ${\sigma}_{\alpha}{\cal
P}_{{\alpha}{\beta}}$ . ${\cal L}_{\beta}$ and $J_{\beta}$ are
orbital and total angular momenta respectively.

If $f\in{\cal A}=C^{\infty}(S^2)$, it has the partial wave
expansion:
\begin{equation}
f(x)=\sum_{k M}f^k_M <kM| D^{(k)}(g)|k0>
\label{pw}
\end{equation} 
where  $D^{(k)}:g\rightarrow D^{(k)}(g)$ define the
angular momentum $k$ IRR of $SU(2)$ and $g\,\sigma_3\,g^{-1}=\sigma\cdot x$.
The action of  ${\cal L}_\alpha$
on it is specified by:
\begin{equation}
{\cal L}_\alpha<kM| D^{(k)}(g)|k0>=-<kM|J_\alpha^{(k)}\, D^{(k)}(g)|k0>
\end{equation}
 where $J_{\alpha}^{(k)}$ are angular momentum $k$
 $SU(2)-$generators.

${\cal D}$ acts on ${\cal A}{\otimes}C^2{\equiv}{\cal
A}^2=\{(a_{1/2},a_{-1/2}):a_\lambda{\in}C^{\infty}(S^2)\}$. 
It anticommutes with the chirality operator
\begin{equation}
{\Gamma}={\sigma}\cdot\hat{x}.
\end{equation}
We now find the eigenfunctions of ${\Gamma}$.

Following (\ref{pw}), we can
define a function $x_\alpha$ on $SU(2)$ as follows. For
$g{\in}SU(2)$, $x_\alpha(g)$ is defined by
$g{\sigma}_3g^{+}={\sigma}\cdot x(g)$. 
${\sigma}\cdot\hat{x}$ is now the
chirality operator on ${\cal A}^2$ defined in the following way. 
 The action of
${\sigma}\cdot\hat{x}$ on $D^{(k)}$ is specified by
$[{\sigma}\cdot\hat{x}D^{(1/2)}](g)={\sigma}\cdot x(g)D^{(1/2)}(g)$. We will
henceforth often omit $g$ in writing $x(g)$. Since $D^{(1/2)}(g)=g$,
it follows that helicity ${\pm}1$ eigenfunctions of
${\sigma}\cdot\hat{x}=D^{(1/2)}\sigma_3 {D^{(1/2)}}^{-1}$ are
\begin{equation}
D^{(1/2)}_{.,{\pm}1/2}=\left(D^{(1/2)}_{1/2,{\pm}1/2},D^{(1/2)}_{-1/2,{\pm}1/2}
\right).
\end{equation} 
Here, $D^{(1/2)}_{.,{\pm}1/2}\equiv
\hat{g}_{.,{\pm}1/2}$, $\hat{g}_{ij}$ being functions on $SU(2)$:
$\hat{g}_{ij}(g)=g_{ij}$. They have the equivariance property
\begin{equation}
D_{.,{\pm}1/2}^{(1/2)}(ge^{i{\sigma}_3{\theta}})=D^{(1/2)}_{.,{\pm}1/2}(g)e^{{\pm}i{\theta}}.
\label{s51}
\end{equation}

Unlike (\ref{s51}), elements of ${\cal A}^2$ and hence too its
chirality ${\pm}1$ subspaces
$\frac{1{\pm}{\sigma}.\hat{x}}{2}{\cal A}^2$ are invariant under
$g\rightarrow ge^{i{\sigma}_3{\theta}}$ . The expansion of
elements of these subspaces using the above $D$'s must thus have
another $D$ in each term transforming with the opposite phase to
that in (\ref{s51}). Accounting for this fact, we can write
for $a\in {A}^2$,
\begin{eqnarray}
a&=&a^++a^-,\nonumber\\
a^{{\pm}}&=&(a^{\pm}_{1/2},a^{\pm}_{-1/2}){\in}\frac{1{\pm}{\sigma}.
\hat{x}}{2}{\cal
A}^2,\nonumber\\
a^{\pm}_{\lambda}&=&\sum_{n,j}{\xi}_n^{j\pm}D^{(j)}_{n,{\mp}1/2}
D^{(1/2)}_{\lambda,{\pm}1/2},\ \ \ {\xi}^{j\pm}_{n}{\in}{\mathbb C}.
\label{s512}
\end{eqnarray}

Now orbital angular momentum ${\cal L}_{\beta}$ is not defined on
the individual factors in (\ref{s512}). We must lift it to the operator
${\cal J}_{\beta}^L$ which acts on $D^{(j)}$ and
$D^{(\frac{1}{2})}$ in such a manner that
$\hat{x}$ transforms like a vector; ${\cal J}_{\beta}^L$
are $SU(2)$ generators acting by left translation:
\begin{eqnarray}
[e^{i{\theta}_{\beta}{\cal J}_{\beta}^L}
D^{(\frac{1}{2})}_{ij}](g)&=&D^{(\frac{1}{2})}_{ij}
(e^{-i{\theta}_{\beta}\frac{{\sigma}_{\beta}}{2}}g)\nonumber\\
&=&[e^{-i{\theta}_{\beta}\frac{{\sigma}_{\beta}}{2}}g]_{ij}.
\label{s52}
\end{eqnarray}
We now reinterpret $J_\beta$ as
\begin{equation}
J_{\beta}={\cal J}_{\beta}^L+\frac{{\sigma}_{\beta}}{2}.
\end{equation}
Because of the transformation rule (\ref{s52}), we find,
\begin{equation}
J_{\beta}D^{(1/2)}_{.,{\pm}}=0.
\end{equation}
So
\begin{equation}
{\cal D}D^{(j)}_{n,{\mp}1/2}D^{(1/2)}_{.,{\pm}1/2}=[{\cal
J}_\beta^L D_{n,{\mp}1/2}^{(j)}][{\sigma}_{\alpha}{\cal
P}_{{\alpha}{\beta}}D^{(1/2)}_{.,{\pm}1/2}].
\label{s53}
\end{equation}
Further simplification can be achieved by writing
\begin{eqnarray}
{\sigma}_{\alpha}{\cal P}_{{\alpha}{\beta}}
=[\frac{1}{2}\sigma\cdot\hat x,[\frac{1}{2}\sigma\cdot\hat x,\sigma_\beta]]
={D^{(\frac{1}{2})}}[\frac{1}{2}{\sigma}_3,[\frac{1}{2}{\sigma}_3,{\sigma}_{\alpha}]]
{D^{(\frac{1}{2})-1}}D^{(1)}_{{\beta}{\alpha}}.
\end{eqnarray}
and noticing that
$-D^{(1)}_{{\beta}{\alpha}}\,{\cal J}_{\beta}^L ={\cal
J}_{\alpha}^{R}$ are $SU(2)$ generators acting in the right of $g$, 
\begin{equation}
[e^{i{\theta}_{\alpha}{\cal
J}_{\alpha}^R}f](g)=f[ge^{i{\theta}_{\alpha}\frac{{\sigma}_{\alpha}}{2}}]
\end{equation}
($f:SU(2){\rightarrow}{\mathbb C})$ being a function on $SU(2)$).
Putting this in (\ref{s53}),
\begin{equation}
{\cal D}
D_{n,\mp 1/2}^{(j)}D_{.,\pm 1/2}^{(1/2)}=
-[D^{(j)}_{n,n'}(J^{(j)}_\alpha)_{n',\mp 1/2}]D^{(1/2)}_{.,\mu'}
[\frac{1}{2}\sigma_3,[\frac{1}{2}\sigma_3,\sigma_\alpha]]_{\mu',\pm 1/2}
\end{equation}
The summation on ${\alpha}$ can be
restricted to ${\pm}$ corresponding to raising and lowering 
operators as
the ${\alpha}=3$ term vanishes in the last factor.

It follows that
\begin{eqnarray}
\left({\cal D}a\right)_\lambda =&-&\left\{\sum_{j,n}
 {\xi}^{j+}_n\,D^{(j)}_{n,1/2}
(J_{+}^{(j)})_{1/2,-1/2}\,D^{(\frac{1}{2})}_{\lambda,-1/2}+\right.\nn\\
&&\left. +\sum_{j,n} {\xi}^{j-}_n\,D^{(j)}_{n,-1/2}
(J_{-}^{(j)})_{-1/2,+1/2}\,D^{(\frac{1}{2})}_{\lambda,+1/2}\right\},
\label{afters53}
\end{eqnarray}
%
which can also be written in the ``Dirac-K{\"a}hler'' form
\cite{ref11}
\begin{equation}
{\cal D}\left[\begin{array}{c}
              \sum\xi^{j+}_n D^{(j)}_{n,-1/2} \\
              \sum\xi^{j-}_n D^{(j)}_{n,+1/2}
                \end{array}
         \right]=\!\!-\!\!
                   \left[ \begin{array}{cc}
                     0  & (J^{(j)}_{-})_{-1/2,+1/2}\\
                     (J^{(j)}_{+})_{+1/2,-1/2}  &
                     0
                           \end{array}
                    \right]\left[ 
                            \begin{array}{c}
                  \sum\xi^{j+}_n D^{(j)}_{n,+1/2}  \\
                  \sum\xi^{j-}_n D^{(j)}_{n,-1/2}
                              \end{array}
                            \right]
\label{s54}
\end{equation}
The eigenvalues of ${\cal D}$ are $\epsilon (j+\frac{1}{2})$ with 
$\epsilon=\pm 1$, each with degeneracy $(2j+1)$,
while the corresponding eigenfunctions  have 
$\xi_n^{j\epsilon\pm}=c^{j\epsilon}_n\langle j,\mp\frac{1}{2};\frac{1}{2},
\pm\frac{1}{2}|j-\frac{\epsilon}{2},0\rangle$. Explicitly,
\begin{equation}
(a^{j\epsilon})_\lambda=\sum_n c^{j\epsilon}_n\langle j,n;\frac{1}{2},\lambda|
D(g)|j-\frac{\epsilon}{2},0\rangle
\end{equation}
where additional superscripts  have been added to the 
eigenfunction.

\section{The Dirac Operator on $\CP2$}
\setcounter{equation}{0}


$\CP2$ is not spin, but spin$_c$ \cite{ref11,ref19}. This fact introduces
serious differences between the $\CP2$ Dirac operator and the
Dirac operator for a spin manifold such as ${\mathbb C}P^1$ discussed last.

The $\CP2$ Dirac operator and its fuzzy version have been treated
in \cite{ref11}. Here we develop an alternative approach which seems
capable of generalisation to other coset spaces.

Elsewhere \cite{ref6}, we plan to treat tensor 
analysis on ${\mathbb C}P^N$ and its
fuzzy versions in the language of projective modules. Here we
will summarize just some points relevant for us. The next
section will give their fuzzy versions.

\subsection{The Projective Module For Tangent Bundle and its Complex Structure}

The generators $ad\,{\lambda}_i$ in the adjoint
representation $Ad:g\rightarrow Ad\,g$
of $SU(3)=\{g\}$ have matrix elements
$(ad{\lambda}_i)_{jk}=-2if_{ijk}$ where
$f_{ijk}$ is totally antisymmetric. We have
the identity
$[{\lambda}_i,{\lambda}_j]=
2if_{ijk}{\lambda}_k$
and a similar relation for $ad\,{\lambda}_i$. 
As hypercharge commutes with
itself and isospin generators, it follows that $f_{8ij}=0$ if
$i$ or $j=1,2,3$ or $8$. Thus the tangent vectors to $\CP2$ at
${\xi}^{0}=(0,..0,1)$, or equivalently at 
$\lambda_8=\lambda_i\xi^0_i$, 
are $ad{\lambda}_j$, $j=4,5,6,7$. 
The directions $ad{\lambda}_j$, $j=1,2,3,8$ are normal.
At any other point ${\xi_i\lambda_i}=g\lambda_8g^{-1}{\in}\CP2$, 
the normals accordingly are  
 $Ad\,g(ad\,\lambda_j)\, Ad\,g^{-1}:=\xi^{(j)}_iad{\lambda}_i$, $j=1,2,3,8$
where $\xi_i^{(8)}=\xi_i$. 
That means that $f_{ikl}\xi_k\xi^{(j)}_l=0$
The four orthogonal directions
$Ad\,g(ad\,\lambda_j)\, Ad\,g^{-1}\,\,\,(j=4,5,6,7)$
in the trace norm span the tangent space.

The eigenvalues of $ad\,Y\,\,(Y=\frac{1}{\sqrt{3}}{\lambda}_{8})$, and hence 
also of $\frac{1}{\sqrt{3}}{\xi}_iad{\lambda}_i$ 
are ${\pm}1,0$ corresponding to the 
mesons $K$, $\bar{K}$,${\eta}^{0}$ and $\vec{\pi}$ in the 
flavor octet terminology. If ${\chi}^{(+)}$ is an
eigenvector for eigenvalue $+1$, $\frac{1}{\sqrt{3}}({\xi}_i ad{\lambda}_i)
{\chi}^{(+)}={\chi}^{(+)}$, then $\xi^{(j)}_i \chi_i^{(+)}=
\frac{1}{\sqrt{3}}{\xi}_k^{(j)}(\xi_l^{(8)}ad{\lambda}_l)_{ki}
{\chi}_i^{(+)} =0$ from above where $j=1,2,3,8$. Hence ${\chi}^{(+)}$ 
is a tangent at ${\xi}$. 
So is ${\chi}^{(-)}$ for eigenvalue $-1$. Hence ${\chi}^{(\pm)}$ span the 
tangent space and the null space of ${\xi}_iad{\lambda}_i$ 
spans the space of normals.

We can now present sections of the tangent bundle $T\CP2$ as a 
projective module. 
Let ${\cal A}^8={\cal A}{\otimes}C^8=\{({\hat{\xi}}_1,...,{\hat{\xi}}_8)\}$.
\begin{equation}
\frac{1}{3}({\hat{\xi}}_iad{\lambda}_i)^2={\cal P}
\label{prj}
\end{equation}
is a projector and ${\cal P}{\cal A}^8$ is seen to
consist of the sections of tangent bundle from the above remarks.

The complex structure on $\CP2$ can be thought of as a splitting
of the tangent space $T_{{\xi}}\CP2$ as the direct sum
$T_{{\xi}}^{(+)}\CP2+T_{{\xi}}^{(-)}\CP2$ for all
${\xi}{\in}\CP2$ in a smooth manner . The tensor ${\cal J}$ of
complex analysis at ${\xi}$ is then ${{\pm}i}$ on 
$T_{{\xi}}^{(\pm)}\CP2$.

In the language of projective modules , we must thus split ${\cal
P}$ as the sum of two orthogonal projectors ${\cal P}^{(\pm)}$.
The tensor ${\cal J}$ is ${\pm}i$ on ${\cal P}^{(\pm)}{\cal A}^8$, 
that is, ${\cal J}=i({\cal P}^{(+)}-{\cal P}^{(-)})$. Hence
also ${\cal J}{\cal P}={\cal P}{\cal J}={\cal J}$.

$SU(3)-$covariance suggests the choice of ${\cal P}^{(\pm)}{\cal A}^8$ 
as eigenspaces of $\frac{1}{\sqrt 3}\hat{\xi}_i ad{\lambda}_i$ 
for eigenvalues ${\pm}1$.
Hence
\begin{equation}
{\cal P}^{(\pm)}=\frac{1}{2\sqrt{3}}{\hat{\xi}}_iad{\lambda}_i
(\frac{1}{\sqrt 3}\hat{\xi}_jad{\lambda}_j{\pm}1).
\end{equation}
As $ad{\lambda}_{\alpha}$ is antisymmetric, we have that
\begin{equation}
{\cal P}^{(+)T}={\cal P}^{(-)},{\cal J}^{T}=-{\cal J}.
\end{equation}

From ${\cal J}$, we can also write the Levi-Civita symbol in an
$SU(3)-$covariant way. It is
\begin{equation}
{\epsilon}_{{\alpha}{\beta}{\gamma}{\delta}}=3{\cal J}
_{[{\alpha}{\beta}}{\cal
J}_{{\gamma}{\delta}]},~[~]:\ \ {\rm antisymmetrisation}.
\end{equation}

\subsection{The Gamma Matrices}


Since $\CP2$ is a submanifold of $R^8$ , it is natural to start from the 
Clifford algebra on $R^8$. Let its basis be the $16{\times}16$ matrices 
$\hat{\gamma}_i$ 
$(i=1,2,...8)$ with the relations
\begin{equation}
\{\hat{\gamma}_i,\hat{\gamma}_j\}=2{\delta}_{ij},
\ \ \ \hat{\gamma}_i^{\dagger}=\hat{\gamma}_i
\end{equation}
The ${\gamma}-$ matrices which will occur in the Dirac operator are not 
these,
rather they will be $16{\times}16$ ${\gamma}-$matrices ${\gamma}_{\mu}$ 
with the same relations
\begin{equation}
\{{\gamma}_i,{\gamma}_j\}=2{\delta}_{ij},
\ \ \ {\gamma}_i^{\dagger}={\gamma}_i
\end{equation}
but which act by left multiplication on the algebra generated by
$\hat{\gamma}_i$, that is on the algebra $Mat_{16}$ of
$16{\times}16$ matrices $M$. Thus
\begin{equation}
{\gamma}_iM={\hat{\gamma}}_iM.
\end{equation}
The matrices of $Mat_{16}$ have a scalar 
product $(M,N)={\rm Tr}\,(M^{\dagger}N)$ 
for which ${\gamma}_i^{\dagger}={\gamma}_i$.

The $\CP2$ ${\gamma}'s$ are the tangent projections 
${\gamma}_i{\cal P}_{ij}$.
There are only four of them at each ${\xi}$ which are linearly independent.
We have to find a four-dimension subspace of $Mat_{16}$ 
at each ${\xi}$ on which they can act. If we fail in that, we will end 
up with more than one fermion.

We first find this subspace at ${\xi}^0$. At ${\xi}^0$, define the fermionic 
creation-annihilation operators
\begin{eqnarray}
\hat{a}^{\dagger}_{1}&=&\frac{1}{2}({\hat{\gamma}}_4+i{\hat{\gamma}}_5), \hat{a}_{1}=\frac{1}{2}({\hat{\gamma}}_4-i{\hat{\gamma}}_5),\nonumber\\
\hat{a}^{\dagger}_{2}&=&\frac{1}{2}({\hat{\gamma}}_6+i{\hat{\gamma}}_7),\hat{a}_{2}=\frac{1}{2}({\hat{\gamma}}_6-i{\hat{\gamma}}_7).
\end{eqnarray}
$\hat{a}^{\dagger}_{\alpha}$ transform as $(K^{+},K^{0})$, 
$\hat{a}_{\alpha}$ as $(K^{-},\bar{K}^0)$. Let
\begin{equation}
|0\rangle={\hat{a}}_1{\hat{a}}_2,\ \ 
|\alpha\rangle=\hat{a}_{\alpha}^{\dagger}|0\rangle\ ({\alpha}=1,2),\ \ 
|3\rangle=\hat{a}^{\dagger}_{1}\hat{a}^{\dagger}_{2}|0\rangle.
\end{equation}
They span a $4-$dimensional space. ${\gamma}_{\alpha}$ 
$(4{\leq}{\alpha}{\leq}7)$ act irreducibly on this space. If
\begin{equation}
a_1^{\dagger}=\frac{1}{2}({\gamma}_4+i{\gamma}_5),a_2^{\dagger}=
\frac{1}{2}({\gamma}_6+i{\gamma}_7)
\end{equation}
and their adjoints define their creation-annihilation operators, 
$|0\rangle$ is their vacuum state.

For an appropriate subspace at other points of $\CP2$, we use the fact 
that $SU(3)$ acts transitively on $\CP2$. 
Thus we can regard ${\xi}{\in}\CP2$ as a function on $SU(3)$ 
with value $\xi(g)$ at $g$
via the relation $g{\lambda}_8g^{-1}={\lambda}_i{\xi}_{
i(g)}$. Then ${\xi}^0={\xi}(e)$, $e={\rm identity}$.

Now the algebra of $SU(3)$ can be realised using ${{\gamma}}_i$, 
the generators being
\begin{equation}
t_i^c=\frac{1}{4i}f_{ijk}\gamma_j\gamma_k
\end{equation}
Their $16-$dimensional representation can be split into 
$\underline{8}{\oplus}\underline{8}$ using the projectors 
$P_{\pm}=\frac{1{\pm}{\hat{\gamma}}_{9}}{2}$ , ${\hat{\gamma}}_9=
{\hat{\gamma}}_1{\hat{\gamma}}_2...{\hat{\gamma}}_8$.
The ${{\gamma}}_i$
transform as an $\underline{8}$ under the action
${{\gamma}}_i{\rightarrow}[t_j^c,{{\gamma}}_i]$ 
by derivation:
\begin{equation}
[t_j^c,{\gamma}_i]:=ad\,t_j{\gamma}_i=if_{jil}{{\gamma}}_l.
\label{s55}
\end{equation}
Let $T(g)$ be the image of $g$ in the $SU(3)$ representation given by
(\ref{s55}). 
$T(g)$ can act on $Mat_{16}$ by conjugation according to
$Ad\,T(g)M=T(g)MT(g)^{-1}$. $ad\,{t_i}$ are the infinitesimal
generators for the action $Ad\,T(g)$ of $SU(3)$.

The $4-$dimensional vector space at $g=e$ and its basis can be 
labelled as $V(e)$ and 
$\{|{\nu};e\rangle,{\nu}=0,{\alpha},3:|{\nu};e\rangle=|\nu\rangle\}$.
The vector space and its basis at $g$ are then
\begin{eqnarray}
V(g)&=&Ad\,T(g) V(e)=T(g)V(e)T(g)^{-1},\nonumber\\
|{\nu};g\rangle &=& Ad\,T(g)|\nu;e\rangle=T(g)|\nu;e\rangle T(g)^{-1}.
\end{eqnarray}
It is on this vector space that ${\gamma}_iP_{ij}({\xi}(g))$ acts by 
left-multiplication.

On the vector space $V(e)$, the $U(2)$ subgroup of $SU(3)$ 
acts by conjugation.
From the particle physics interpretation 
of $\hat{a}^{+}_{\alpha}$, we see that $V(e)$ decomposes into the direct sum
\begin{equation}
(I=0,Y=-2){\oplus}(\frac{1}{2},-1){\oplus}(0,0).
\end{equation}

To see the $SU(3)$ representation content of $|\nu,g\rangle$, 
let us first focus on $|0;g\rangle$. $|0;e\rangle\equiv|0\rangle$ is 
bilinear 
and antisymmetric in the ${\gamma}'s$ and has $I=0,Y=-2$. The action 
$T(g)$ preserves the number of ${\gamma}'s$. Thus its $SU(3)-$ orbit
is contained in the vector space spanned by the antisymmetric product 
of two ${\gamma}'s$, that is, 
${\gamma}_{ij}=\frac{1}{2}(\gamma_i\gamma_j-\gamma_j\gamma_i)$.
This vector space transforms as 
${\underline{10}}{\oplus}\bar{{\underline 10}}{\oplus}{\underline{8}}$. 
Only ${\underline{10}}$ 
contains an $I=0,Y=-2$ vector, namely ${\Omega}^{-}$, thus 
$|0;g\rangle{\in}\underline{10}=(N_1=3,N_2=0)$.

A more explicit formula can be written. Let $|(3,0);(I,I_3,Y);e\rangle$ 
be the basis of vectors, which are linear in ${\gamma}_{ij}$ and transforms 
as ${\underline{10}}$. We have: $|(3,0);(0,0,-2);e\rangle\equiv|0;e\rangle$. 
Then
\begin{equation}
|0;g\rangle = Ad\,T(g)\,|0;e\rangle
=|(3,0);(I,I_3,Y);e\rangle D^{(3,0)}_{(I,I_3,Y);(0,0,-2)}(g)
\label{s56}
\end{equation}
where $D^{(3,0)}:g{\rightarrow}D^{(3,0)}(g)$ is the 
IRR $\underline{10}$ of $SU(3)$ and the basis is labelled by $(I,I_3,Y)$.

We can analyse the $SU(3)$ content and write an explicit formula for
every $|\nu;g\rangle$ \cite{ref20}. $|\alpha;e\rangle(\alpha=1,2)$ has
${\gamma}_i$'s and ${\gamma}_{ijk}$'s where we mean by
${\gamma}_{ijk..l}$ the totally antisymmetrised product of
${\gamma}_i$, ${\gamma}_j$, ${\gamma}_k$,
...,${\gamma}_l$. ${\gamma}_i$ transforms as an $\underline{8}$ or
$(N_1=1,N_2=1)$ while ${\gamma}_{ijk}$ transforms as
$\underline{27}{\oplus}\underline{10}{\oplus}\bar{\underline{10}}{\oplus}
\underline{8}{\oplus}\underline{1}$. We can take
linear combination of ${\gamma}_i$ and ${\gamma}_{ijk}$ to form
two new $\underline{8}$'s such that the
$\underline{8}$ part of $|\alpha;e\rangle$ is in a single $\underline{8}$.
Also $|\alpha;e\rangle$ has $I=1/2$,$Y=-1$ and such a vector occurs only
in $\underline{8}$, $\underline{10}$ and $\underline{27}$.
Thus $|\alpha;g\rangle\,\,(\alpha=1,2)$ transforms
as the direct sum $\underline{8}{\oplus}\underline{10}{\oplus}\underline{27}$;
calculating explicitely the coefficients we find:
\begin{eqnarray}
|\alpha;g\rangle&=&\sqrt{\frac{3}{5}}\,|(1,1);(I,I_3,Y);e\rangle
D^{(1,1)}_{(I,I_3,Y),(1/2,(3-2{\alpha})/2,-1)}(g)\nonumber\\
&-&\frac{1}{2}|(3,0);(I,I_3,Y);e\rangle
D^{(3,0)}_{(I,I_3,Y),(1/2,(3-2{\alpha})/2,-1)}(g)\nonumber\\
&+&\sqrt{\frac{3}{20}} |(2,2);(I,I_3,Y);e\rangle
D^{(2,2)}_{(I,I_3,Y),(1/2,(3-2{\alpha})/2,-1)}(g)
\end{eqnarray}

There remains $|3;e\rangle$ with $I=Y=0$. It is a linear combination
of a constant, ${\gamma}_{ij}$, and ${\gamma}_{ijkl}$. i) The
constant part transforms as an $SU(3)-$singlet.
ii) ${\gamma}_{ij}$ was treated above.  iii) ${\gamma}_{ijkl}$ is
$\underline{27}{\oplus}\underline{8}$. 
$U(2)$ singlets with $I=Y=0$ are contained only
in $SU(3)$ singlet $\underline{8}$ and $\underline{27}$ so that $|3;g\rangle$
transforms as $\underline{1}{\oplus}\underline{8}{\oplus}\underline{27}$,
the $\underline{8}$ being a mixture of $\underline{8}'s$ from ${\gamma}_{ij}$, 
${\gamma}_{ijkl}$.
Calculating the coefficients explicitely, we find,
\begin{eqnarray}
|3;g\rangle&=&-\frac{1}{2}\,|0;(0,0,0);e\rangle\nonumber\\
&+&\sqrt{\frac{3}{5}}\,|(1,1);(I,I_3,Y);e\rangle
D^{(1,1)}_{(I,I_3,Y),(0,0,0)}(g)\nonumber\\
&+&\sqrt{\frac{3}{20}}\,|(2,2);(I,I_3,Y);e\rangle
D^{(2,2)}_{(I,I_3,Y),(0,0,0)}(g).
\end{eqnarray}
 

\subsection{The Dirac Operator}


We require of the $\CP2$ Dirac operator ${\cal D}$ that it is linear 
in derivatives and anticommutes with the chirality operator ${\Gamma}$:
\begin{eqnarray}
{\Gamma}:=-\frac{1}{4!}{\epsilon}_{ijkl}\gamma_i\gamma_j\gamma_k\gamma_l.
\end{eqnarray}
At ${\xi}={\xi}^0$, ${\Gamma}={\gamma}_4{\gamma}_5{\gamma}_6{\gamma}_7$ and 
is $+1$ on $|0;e\rangle$ and $|3;e\rangle$, and $-1$ on 
$|\alpha;e\rangle (\alpha=1,2)$. Hence ${\Gamma}=+1$ on 
$|0;g\rangle,|3;g\rangle$ and $-1$ on $|{\alpha};g\rangle$ for all $g$. 
The former have even chirality and the latter have odd chirality.

Now ${\gamma}_iP_{ij}$ anticommutes with ${\Gamma}$, while
the $SU(3)$ generators
\begin{eqnarray}
J_i&=&{\cal L}_i + ad\,{t_i^c},\nn\\
{\cal L}_i&=&-if_{ijk}{\hat\xi}_j\frac{\partial}{\partial
{\hat\xi}_k}
\end{eqnarray}
commute with ${\Gamma}$. Hence
\begin{equation}
{\cal D}={\gamma}_i{\cal P}_{ij}J_j
\label{s5extra1}
\end{equation}
anticommutes with ${\Gamma}$,
\begin{equation}
\{{\Gamma},{\cal D}\}=0,
\end{equation}
and is a good choice for the Dirac operator.

${\cal D}$ acts on ${\cal A}{\otimes}Mat_{16}$. But there are
only four tangent gammas at each ${\xi}(g)$, so we have to
reduce ${\cal A}{\otimes}Mat_{16}$ to 
$V(g)$ (in an appropriate sense)
at each ${\xi}(g)$. We can achieve this reduction as
follows. The functions $\hat{\xi}$ are defined according to
\begin{equation}
\hat{\xi}(g)=T(g)\,t^c_8\,T^{-1}(g).
\label{s5extra}
\end{equation}
where the notation means that $T$ and  $T^{-1}$ are to be evaluated
at $g$.
Hence if
$u{\in}U(2)$, the stability group of 
$t^c_8$, $\hat{\xi}(gu)=\hat{\xi}(g)$.
This means that ${\cal A}{\otimes}Mat_{16}$ consists of 
sections of the trivial $U(2)-$bundle over $\CP2$.
The same is the case for its left- and right- chiral projections
\begin{equation}
{\Psi}_{\pm}=\frac{1{\pm}{\Gamma}}{2}{\cal A}{\otimes}Mat_{16}.
\end{equation}

But that is not the case for $|0;g\rangle$ and $|\alpha;g\rangle$.
Under $g{\rightarrow}g\,u$, $|0;g\rangle$ transforms as an $SU(2)$ singlet 
with $Y=-2$ and $|\alpha;g\rangle$
transforms as an $SU(2)$ doublet with hypercharge $Y=-1$.

Let $\hat g$ denote the matrix of functions on $SU(3)$
with ${\hat g}_{ij}(g)=g_{ij}, g\in SU(3)$. 
(${\hat g}$ is just a simplified notation for $D^{(1,0)}$).
We regard elements of ${\cal A}{\otimes}Mat_{16}$ as functions of ${g}$,
invariant under the substitution $g\rightarrow g\,u$. 
Accordingly, let us also introduce the vectors 
$|a;\hat{g}\rangle;a=0,\alpha,3$ 
which at $g$ are the vectors $|a;\hat{g}(g)\rangle=|a;g\rangle$. 
Note that on a function $f$ on $SU(3)$, the left- 
and right- actions of $h{\in}SU(3)$ are $f{\rightarrow}h^{L,R}f$ where 
$(h^{L}f)(g)=f(h^{-1}g)$ and $(h^Rf)(g)=f(gh)$.

Now consider, in the case of $|0,g\rangle$, the wave functions
$D^{(N_1,N_2)}_{(II_3Y)(0,0,2)}$. They exist only if $N_2=N_1+3$. 
The combination
\begin{equation}
D^{(N,N+3)}_{(I,I_3,Y)(0,0,2)}|0,\hat{g}\rangle
\label{s57}
\end{equation}
is invariant under $g{\rightarrow}gu$ at each $g$ and can form
constituents of a basis for the expansion of functions in ${\cal
A}{\otimes}Mat_{16}$.

The remaining elements of a basis can be found in the same manner,
being
\begin{eqnarray}
&&\frac{1}{\sqrt{2}}D^{(N_1,N_2)}_{(I,I_3,Y)(\frac{1}{2},-m,1)}
|\tilde{m},\hat{g}\rangle;N_2-N_1=0\ \ {\rm or}\ \ 3,\nonumber\\
&&D^{(N,N)}_{(I,I_3,Y)(0,0,0)}|3,\hat{g}\rangle
\label{s58}
\end{eqnarray}
where $\tilde{\frac{1}{2}}$,$-\tilde{\frac{1}{2}}$ stand
for ${\alpha}=1,2$ and $m$ is summed over.

There is a subtlety we encounter at this point. [We also came
across it for $S^2$]. "Orbital" $SU(3)$ momentum ${\cal
L}_{\alpha}$ does not act on the individual factors in (\ref{s57},\ref{s58}),
which are functions on $SU(3)$ and not just $\CP2$. It
is thus necessary to lift them to operators ${\cal\hat J}_i^L$
which act on $\hat{g}$ in such a manner that when (\ref{s5extra}) is used,
$\hat{\xi}$ transform under $SU(3)$ in the way desired:
${\hat{\xi}}{\rightarrow}h^{L}{\hat{\xi}}$. Thus ${\cal
\hat J}_i^L$ are generators of $SU(3)_L$, the left-regular
representation, and the Dirac equation is to be reinterpreted 
with $J_j$ replaced by:
\begin{equation}
J_j={\cal\hat J}_j^L+ ad\,t_j^c. 
\label{s510}
\end{equation}
Restricted to ${\cal A}{\otimes}Mat_{16}$, (\ref{s510}) is the same as
(\ref{s5extra1}).

$|\nu;g\rangle$ is $T(g)|\nu;e\rangle T(g)^{-1}$ so $|\nu;\hat{g}\rangle$ is
$T|\nu;e\rangle T^{-1}$. Now
$(h^{L}T)(g)=T\,(h^{-1}g)$. That is,
$h^{L}(T\,|\nu;e\rangle T^{-1})(g)=T(h^{-1})T(g)
|\nu;e\rangle\, T(g)^{-1}T(h)$, 
which has the infinitesimal form ${\cal
\hat J}_j^L(T\,|\nu;e\rangle T^{-1})(g)=-ad\,t_j^c\,(T(g)
|\nu;e\rangle T(g)^{-1})$. 
We conclude that
\begin{equation}
J_j|\nu,\hat{g}\rangle = 0.
\end{equation}

The expression for ${\cal P}$ is in (\ref{prj}). Using commutation 
relations, 
we can write
\begin{equation}
\gamma_j{\cal P}_{ji}= \frac{4}{3}[t_k^c{\hat{\xi}}_k,
[t_j^c{\hat{\xi}}_j,{\gamma}_i]].
\end{equation}
Now
$Ad\,T\,{\gamma}_j=T\,{\gamma}_jT^{-1}={\gamma}_kAd\,\hat{g}_{kj}$
where $Ad\,\hat{g}(g)=Ad\,g$ represents $g$ in the octet
representation, it is real and orthogonal. Hence
\begin{equation}
\gamma_j{\cal P}_{ji}= \frac{4}{3}\{[Ad\,T][t^c_{8},[t^c_{8},{\gamma}_j]]
[Ad\,T^{-1}]\}{Ad\,{\hat{g}}}_{ij}.
\end{equation}
Since $|\nu,\hat{g}\rangle=Ad\,T|\nu;{e}\rangle$,
$Ad\,T^{-1}|\nu;\hat{g}\rangle=|\nu;{e}\rangle$;
$[t^c_8,[t^c_8,\gamma_i]]$ consists only of tangent space
${\gamma}'s$ at $e$. The action of $\{.\}$ on $|\nu;\hat{g}\rangle$ is thus
\begin{equation}
Ad\,T[t^c_8,[t^c_8,\gamma_i]]Ad\,T^{-1}|\nu;\hat{g}\rangle
=Ad\,T\{[t^c_8,[t^c_8,{\gamma}_i]|\nu;{e}\rangle\}.
\end{equation}

The action of ${\cal D}$ on typical basis vectors like (\ref{s57})
follows:
\begin{eqnarray}
{\cal D}D^{(N,N+3)}_{(I,I_3,Y)(0,0,2)}|0;\hat{g}\rangle= \frac{4}{3}
\{Ad{\hat{g}}_{ij}{\cal
J}_iD^{(N,N+3)}_{(I,I_3,Y)(0,0,2)}\}Ad\,T[t^c_8,
[t^c_8,{\gamma}_j]]|0;{e}\rangle.
\end{eqnarray}

The term in braces also has a considerable simplification. Since
$(h^{L}f)(g)=f(h^{-1}g)$ $=f(g(g^{-1}h^{-1}g))=[({\hat g}^{-1}{\hat h}^{-1}
{\hat g})^{R}f](g)$,
$-Ad{\hat{g}}_{ij}{\cal J}_i^L$ are the
generators ${\cal J}_{\lambda}^{R}$ for the $SU(3)$ acting on the
right of $g$, they have the standard commutation relations
$[{\cal J}^{R}_i,{\cal
J}_j^{R}]=if_{ijk}{\cal J}_k^{R}$. 
We thus find that
\begin{eqnarray}
{\cal D}D^{(N,N+3)}_{(I,I_3,Y)(0,0,2)}|0;\hat{g}\rangle=
- \frac{4}{3}\left\{{\cal J}_i^{R}
D^{(N,N+3)}_{(I,I_3,Y)(0,0,2)}\right\}
\,Ad\,T\,[t^c_8,[t^c_8,{\gamma}_i]]|0;{e}\rangle.
\label{s511}
\end{eqnarray}

The general wave function for even and odd chiralities can be
written as
\begin{eqnarray}
{\xi}^{(i)}_{j}D^{(i)}_{jj^{'}}|j^{''};\hat{g}\rangle;\ \ 
j^{'}=(0,0,2),(0,0,0);\ \  i=N_1,N_2\ , \nn\\ 
N_2-N_1=3 \ \ {\rm if} \ \  j^{'}=(0,0,2) 
\ \ {\rm and}\ \  N_2=N_1 \ \ {\rm if}\ \ j^{'}=(0,0,0)\ , 
\end{eqnarray}
\begin{equation}
{\eta}^{(i)}_{b}D^{(i)}_{bb^{'}}|b^{''};\hat{g}\rangle;\ \ 
b^{'}=(1/2,(3-2\alpha)/2,1);\ \  i=N_1,N_2,\ \ N_1-N_2=0 ~mod\,\,3.
\end{equation}
Here $j^{''}$, $b^{''}$ are the state vectors for ${\gamma}$'s
pairing with $j^{'}$, $b^{'}$ as in (\ref{s57}) and ${\xi}^{(i)}_{j}$,
${\eta}^{(i)}_{b}{\in}{\mathbb C}$ and repeated indices are summed.
Since $\gamma_j{\cal P}_{ji}$ anticommutes with $\Gamma$, we can
represent the effect of ${\cal D}$  on wave functions in terms of
the off-diagonal block
\begin{equation}
\left( \begin{array}{cc}
                                      0&d \\
                                     d^{+}&0
                                    \end{array}
                                     \right),
\label{offdiagform1}
\end{equation}
acting on
\begin{equation}
                                     \left( \begin{array}{c}
                                      {\xi}^{(i)}_{j}D^{(i)}_{jj^{'}}\\
                                       {\eta}^{(i)}_{b}D^{(i)}_{bb^{''}}
                                     \end{array}
                                     \right)
\label{offdiagform2}
\end{equation}
The result is the equation of \cite{ref11} for $m=0$. Ref. \cite{ref11} 
has also found the spectrum of ${\cal D}$.

The zero modes of ${\cal D}$ can be easily worked out from (\ref{s511}).
When $j^{'}=(0,0,0)$, $i$ can be $(0,0)$ [but not otherwise],
and in that case, $D^{(0,0)}$ is a constant and is annihilated
by ${\cal \hat J}^{R}_i$. Hence the index of ${\cal D}$ is
$1$ and the zero mode has even chirality, consistently with \cite{ref11}.

\section{ Fuzzification }
\setcounter{equation}{0}


${\cal D}$ acts on a subspace of
${\cal A}{\otimes}Mat_{16}$. We can thus conceive of a fuzzy Dirac
operator $D$ which acts on a subspace of $A{\otimes}Mat_{16}$,
$A$ being obtained from ${\cal A}$ by restricting ``orbital''
$SU(3)$ IRR's to $(n,n)$, $n{\leq}N$. $D$ is then obtained from
${\cal D}$ by projection to this subspace. ${\cal D}$ commutes
only with the total $SU(3)$ Casimir $J_{i}^2$ and not with
orbital $SU(3)$ Casimir ${\cal L}_{i}^2$. This causes edge
effects distorting the spectrum of $D$ for those states having
$(n,n)$ near $(N,N)$ which ${\cal D}$ mixes with $(n^{'},n^{'})$, 
$n^{'}{\geq}N$. This particular edge phenomenon does not occur
for $S^2={\mathbb C}P^1$ where orbital angular momentum ${\cal
L}^2_{\alpha}$ commutes with the Dirac operator. A way to
eliminate such problems is suggested by the work of 
[6-9]:
We
introduce the cut-off not on the orbital Casimir, but on the
{\it total} Casimir, retaining all states upto the cut-off. That
seems the best strategy as it will give a fuzzy Dirac operator
$D$ with a spectrum exactly that of the continuum operator ${\cal
D}$ upto the cut-off point, and which has chirality
(chirality ${\Gamma}$ of ${\cal D}$ commutes with $J^2_{i}$)
and no fermion doubling. This approach is the same as the method
adopted for $S^2$ in [6-9]. For $S^2$, the edge effect turned up as the
absence of the ${-}E$ eigenvalue subspace for the maximum
total angular momentum when the cut-off is introduced in orbital
angular momentum, and attendant problems with chirality.

$D$ being just a restriction of ${\cal D}$, we can continue to use
(\ref{s5extra1}) in calculation, just remembering the truncation of the
spectrum. That means that the analysis in Section 6 can be used
intact. In the final expressions like (\ref{offdiagform1},\ref{offdiagform2}), 
$i$ labels the IRR and the
Dirac operator acts in subspace with fixed $i$. So the cut-off
can be introduced on $i=N_1,N_2$.

\subsection{Coherent States and Star Products}
These have been treated
in \cite{ref21,ref13,ref6}. Here we summarize the
main points so that we can outline the relation of wave functions
like (\ref{s57}) and those based on matrices for fuzzy physics.

{i) {\it The Case of} $S^2\simeq{\mathbb C}P^1$}

Let us first consider $S^2={\mathbb C}P^1$ and its fuzzy versions. The
algebra $A$ is $Mat_{2l+1}$. $SU(2)$ acts on $A$ on left and right
with generators $L_i^{L}$ and $-L_i^{R}$, and orbital angular
momentum is ${\cal L}_{i}=L_i^L-L_i^R$. The spectrum of ${\cal
L}^2$ is $K(K+1)$, $K=0,1,..,2l$. We can find a basis of
matrices $T^{K}_{M}$ diagonal in ${\cal L}^2$ and ${\cal
L}_3$(with eigenvalue $M$) and standard matrix elements for ${\cal
L}_i$. $A$ acts on a $(2l+1)-$dimensional vector space with the
familiar basis $|l,m\rangle$. $T^{K}_{M}$ are orthogonal, $K(K+1)$
and $M$ being eigenvalues of ${\cal L}^2$ and ${\cal L}_3$:
\begin{equation}
(T^{K}_{M},T^{K^{'}}_{M^{'}}):=\,Tr\,T^{K\dagger}_{M}T^{K^{'}}_{M^{'}}=
~{\rm constant}\times{\delta}_{KK^{'}}{\delta}_{MM^{'}}.
\end{equation}

The above suggests that there is a way to regard $A$ as
``functions'' on $S^2$ with angular momenta cut-off at $2l$. Such
functions are also represented by the linear span of spherical
harmonics $Y_{KM}$, $K{\leq}2l$. We want to clarify the
relation of $Y_{KM}$'s to the matrices $T^{K}_{M}$ in $A$.

Towards this end, let us introduce the coherent states
\begin{equation}
|g;l\rangle=U^{(l)}(g)|l,l\rangle
\end{equation}
induced from the highest weight vector $|l;l\rangle$.
$g{\rightarrow}U^{(K)}(g)$ is the angular momentum $K$ IRR of
$SU(2)$. Note the identity
\begin{equation}
|ge^{i{{\sigma}_3\over 2}{\theta}};l\rangle=e^{il{\theta}}|g;l\rangle.
\end{equation}

It is a theorem \cite{ref21} that the diagonal matrix elements 
$\langle g;l|a|g;l\rangle$
completely determine the operator $a$. Further
$\langle  ge^{i{{\sigma}_3\over 2}{\theta}};l|a|ge^{i{{\sigma}_3\over 2}{\theta}}
;l\rangle
=\langle g;l|a|g;l\rangle$
so that $\langle g|a|g\rangle$ depends only on
\begin{equation}
g{\sigma}_3g^{+}={\sigma}\cdot x,\ \ \sum x^2_{i}=1 ; x{\in}S^2.
\end{equation}
In this way, we have the map $A{\rightarrow}C^{\infty}(S^2)$,
$a{\rightarrow}{\tilde a}$; where 
${\tilde a}(x)=\langle g|a|g\rangle$. In this
map, the image of $T^{K}_{M}$ is $Y_{KM}$ after a phase choice:
\begin{equation}
\langle g;l|T^{K}_{M}|g;l\rangle =Y_{KM}(x).
\label{phasechoice}
\end{equation}
For, under $g{\rightarrow}hg$ , $x{\rightarrow}R(h)x$ where
$h{\rightarrow}R(h)$ is the $SU(2)$ vector representation.
Under this transformation, since
\begin{equation}
Y_{KM}(R(h)x)=D^{(K)}(h)_{MM^{'}}Y_{KM^{'}}(x)
\end{equation}
and
\begin{equation}
T^{K}_{M}{\longrightarrow}U^{}(h)^{-1}T^{K}_{M}U(h)=
D^{(K)}(h)^{}_{MM^{'}}T^{K}_{M^{'}},
\end{equation}
where $h{\longrightarrow}D^{(K)}(h)$ is the angular momentum $K$
IRR of $SU(2)$  in a matrix representation, we have the
proportionality of the two sides. (\ref{phasechoice}) and phase conventions
fix the constant of proportionality.

The map $T^{K}_{M}{\rightarrow}Y_{KM}$ is an isomorphism at the
level of vector spaces. It can be extended to the noncommutative
algebra $A$ by defining a new product on $Y_{KM}$'s, the star
product. Thus consider $\langle g;l|T^{K}_{M}T^{L}_{N}|g;l\rangle$. The
functions $Y_{KM}$ and $Y_{LN}$ completely determine $T^{K}_{M}$
and $T^{L}_{N}$, and for that reason also this matrix element.
Hence it is the value of a function $Y_{KM}*Y_{LN}$, linear in
each factor, at $x$:
\begin{equation}
\langle g;l|T^{K}_{M}T^{L}_{N}|g;l\rangle = [Y_{KM}*Y_{LN}](x).
\end{equation}
The product $*$ here, the star product, extends by linearity to all
functions with angular momenta ${\leq}2l$. The resultant algebra
is isomorphic to the algebra $A$.

The explicit formula for $*$ has been found by Pre\v{s}najder
\cite{ref13} (see
also \cite{ref6}). The image of ${\cal L}_{\alpha}a$ is just
$-i(\vec{x}{\wedge}{\vec{\nabla}})_{\alpha}{\tilde a}$. We will use the
same symbol ${\cal L}_\alpha$ to denote
$-i(\vec{x}{\wedge}{\vec{\nabla}})_\alpha$ derivation. The $*$
product is covariant under the $SU(2)$ action in the sense that
\begin{equation}
{\cal L}_\alpha({\tilde a}*{\tilde b})=({\cal
L}_\alpha{\tilde a})*{\tilde b}+{\tilde a}*({\cal L}_{\alpha}{\tilde b}).
\end{equation}
It depends on $l$ and approaches the commutative product of
$C^{\infty}(S^2)$ as $l{\longrightarrow}{\infty}$.

Coherent states thus give an intuitive handle on the matrix
representation of functions. But on $S^2$, we also have
monopole bundles. Sections of these bundles for Chern  class $n$
are spanned by the rotation matrices $D^{(j)}_{mn}$, $j{\geq}|n|$.
They have the equivariance property
\begin{equation}
D^{(j)}_{mn}(ge^{i{{\sigma}_3\over 2}{\theta}})=D^{(j)}_{mn}(g)e^{in{\theta}}.
\end{equation}
How do we represent them by matrices?

In the first instance, let $n{\geq}0$ and consider the coherent
states 
\begin{eqnarray}
|g;l+n\rangle &=&U^{(l+n)}(g)|l+n,l+n\rangle\nonumber\\
|g;l\rangle&=&U^{(l)}(g)|l,l\rangle.
\end{eqnarray}
They span vector spaces $V_{l+n}$ and $V_{l}$. We can consider
the linear operators $Hom(V_{l+n},V_l)$ from $V_{l+n}$ to $V_{l}$. 
They are $[2l+1]{\times}[2(l+n)+1]$ matrices in a basis of
$V_{l+n}$ and $V_l$, and have $U^{(l)}(g)$ acting on their
left(with generators $L^{L}_i$) and $U^{(l+n)}(g)$ acting on
their right (with generators $-L^{R}_i$). We can decompose
$Hom(V_{l+n},V_l)$ under the ``orbital'' angular momentum group
$U^{(l)}{\otimes}U^{(l+n)}$ (with generators ${\cal
L}_\alpha=L^{L}_{\alpha}-L^{R}_\alpha$) into the direct sum
${\oplus}_{K=n}^{2l+n}(K)$ with the IRR $K$ having the basis
$T^{K}_{M}$, with ${\cal L}_{3}T^{K}_{M}=MT^{K}_{M}$. As
before, we choose $T^{K}_{M}$ so that ${\cal L}_i$ follow 
standard phase conventions. $T^{K}_{M}$ are orthogonal
\begin{equation}
Tr(T^{K^{'}}_{M^{'}})^{\dagger}T^{K}_{M}=
~{\rm constant}\times{\delta}_{K^{'}K}{\delta}_{M^{'}M}.
\end{equation}

Now consider
\begin{equation}
\langle l;g|T^{K}_{M}|g;l+n\rangle.
\end{equation}
It transforms in precisely the same manner as $D^{(K)}_{Mn}(g)$ under 
$g{\rightarrow}hg$ and $g{\rightarrow}ge^{i{{\sigma}_3\over 2}{\theta}/2}$ 
and hence after an overall normalisation,
\begin{equation}
\langle l;g|T^{K}_{M}|g;l+n\rangle=D^{(K)}_{Mn}(g).
\end{equation}
Thus $Hom(V_{l+n},V_l)$ are fuzzy versions of sections of vector bundles 
for Chern class $n{\geq}0$. For $n<0$, they are 
similarly $Hom(V_l,V_{l+|n|})$.
This result is due to \cite{ref22} (see also 
[6-9,14]
An explicit formulae for the fuzzy version of rotation matrices
can be found in \cite{ref13}.

It is interesting that Chern class has a clear meaning even in this 
matrix model: It is $|V|-|W|$ for $Hom(V,W)$, where $|V|$ and $|W|$ are 
dimensions of $V$ and $W$.

There are two (inequivalent) fuzzy algebras acting 
on $Hom(V,W)$. $Mat_{|V|}:=A_{|V|}$ acts on the 
right and $Mat_{|W|}:=A_{|W|}$ acts on the 
left, where now a subscript has been introduced on $A$. These left 
and right actions have their own
 $*$'s, call them
$*_{|V|}$ and $*_{|W|}$: if $a{\in}A_{V}$, $b{\in}A_{W}$ and ${\tilde a}$ and 
${\tilde b}$ 
are the corresponding functions, then
\begin{equation}
aT^{K}_{M}b{\longrightarrow}{\tilde a}*_{|W|}Y_{KM}*_{|V|}{\tilde b}
\end{equation}
under the map of $Hom(V,W)$ to sections of bundles.

There is also a fuzzy analogue for tensor products of bundles. Thus we 
can compose elements of $Hom(V,W)$ and $Hom(W,X)$ to get $Hom(V,X)$
\begin{equation}
Hom(V,X)=Hom(V,W){\otimes}_{A_{|W|}}Hom(W,X).
\end{equation}
Its elements are $ST$, $S{\in}Hom(V,W)$, $T{\in}Hom(W,X)$. 
Its Chern class is $|V|-|X|$. If ${\tilde S}$ and ${\tilde T}$ are the 
representatives of $S$ and $T$ in terms of sections of bundles, 
then $ST{\longrightarrow}{\tilde S}*{\tilde T}$.

Tensor products ${\Gamma}_{1}{\otimes}{\Gamma}_2$ of two vector 
spaces ${\Gamma}_1$ and ${\Gamma}_2$ over an algebra $B$ are defined 
only if ${\Gamma}_1({\Gamma}_2)$ is a right-(left-) $B$-module
\cite{ref23}. 
Hence $Hom(V,W){\otimes}_{A_{|W|}}Hom(W^{'},X)$
is defined only if $W=W^{'}$. So ${\tilde S}*{\tilde T}$ is rather 
different in its properties
from the usual tensor product of bundle sections, in 
particular ${\tilde T}*{\tilde S}$ makes no sense if $V\ne X$. 

We can now comment on the fuzzy form of (\ref{s54}). Elsewhere the Watamuras
\cite{w1,w2} 
and following them, us \cite{ref9,ref12}, 
investigated the Dirac operator as acting on 
$A{\otimes}C^2=A^2$, $A=Mat_{2l+1}$. That led to rather an elaborate 
formalism because of the cut-off in orbital 
angular momentum. So as indicated earlier, it seems more elegant to 
cut-off total angular momentum at some value $j_0$.

We can now argue such a cut-off leads to the formalism of 
[5-9]
and to 
supersymmetry. Thus let $T^{j}_{n+}{\in}Hom(V_{l+1/2},V_l)$ 
with the transformation property
\begin{equation}
U^{(l)}(g)^{\dagger}T^{j}_{n+}U^{(l+1/2)}(g)=D^{(j)}_{nn^{'}}(g)
T^{j}_{n^{'}+}
\end{equation}
[So $j{\leq}2l+1/2$ and $j_0=2l+1/2$].
Then
\begin{equation}
D^j_{n+}(g)=\langle g; l|T^j_{n+}|g;l+1/2 \rangle 
\end{equation}
and an overall constant of proportionality has been set 
equal to 1 by suitably scaling $T^{j}_{n+}$.
The subscript $+$ indicates helicity $-$ [see Eq.(\ref{s512})].

For helicity $+$, but for same $j_0$, we have to consider 
$T^{j}_{n-}{\in}Hom(V_l,V_{l+1/2})$, with
\begin{equation}
U^{(l+1/2)}(g)^{\dagger}T^{j}_{n-}U^{(l)}(g)=D^{(j)}_{nn^{'}}(g)T^{j}_{n^{'}-}.
\end{equation}
This is the formalism of [5-9]
As we have united $V^{(l)}$ and $V^{(l+1/2)}$, 
it is natural to consider $OSp(2,1)$ or 
even $OSp(2,2)$ $SUSY$ as discovered first by Grosse et al. 
in the second paper of \cite{ref15}.
The action of the fuzzy Dirac operator $D$ on $T^{j}_{n{\pm}}$ is merely 
the truncated form of (\ref{s54}):
\begin{equation}
DT_{n{\pm}}=-\left(J^{(j)}_\pm\right)_{\pm\frac{1}{2},\mp\frac{1}{2}}
T^{j}_{n{\mp}}, \ \ j{\leq}2l+1/2.
\end{equation}

Because of the mixing of $l$ and $l+1/2$, we have to reconsider 
the action of the matrix algebra $A$ approximating 
${\cal A}={C}^{\infty}(S^2)$. $Mat_{2l+1}$ acts on 
$T^{j}_{n+}(T^{j}_{n-})$ on the 
left(right) while $Mat_{2l+2}$ acts 
on $T^{j}_{n+}(T^{j}_{n{-}})$ on the right(left). So it is best to 
regard fuzzy functions to act on left(say) of $T^j_{n+}$ and right 
of $T^j_{n-}$ as $Mat_{2l+1}$.
This suggestion is slightly different from that of 
[5-9]
where they regard 
the fuzzy algebra to be $Mat_{2l+1}$ on $T^{j}_{n+}$ 
and $Mat_{2l+2}$ on $T^{j}_{n-}$, both acting on left.
However, our proposal does not generalize to instanton (monopole)
sectors.

We can restore spin parts to fuzzy wave functions. 
The spin wave functions for helicity $\pm$ are $T^{1/2}_{m{\pm}}$. 
So the two components of the total fuzzy wave functions for helicity $\pm$ are
\begin{equation}
\sum {\xi}^{j\pm}_{n} T^{j}_{n{\mp}}T^{1/2}_{\lambda{\pm}},\ \ \ 
{\xi}^{j\pm}_{n}{\in}{\mathbb C}, \lambda=1,2.
\end{equation}
The Dirac operator $D$ is given by the truncated 
version of (\ref{afters53}):
\begin{eqnarray}
&&D_{\lambda\lambda'}\left\{\sum\xi^{j+}_{n}
T^{j}_{n-}T^{1/2}_{\lambda'+}+\sum\xi^{j-}_{n}T^{j}_{n+}T^{1/2}_{\lambda'-}\right\}_{m}
=\nn\\
&&-\left\{\sum\xi^{j+}_{n}T^{j}_{n+}(J^{(j)}_{+})_{+1/2,-1/2}\right\}
\left\{T^{1/2}_{\lambda-}\right\}
-\left\{\sum\xi^{j-}_{n}T^{j}_{n-}(J^{(j)}_{-})_{-1/2,+1/2}\right\}
\left\{T^{1/2}_{\lambda+}\right\},\nonumber\\
&&j{\leq}2l+1/2,
\end{eqnarray}
$J^{(j)}_{\alpha}$ being the angular momentum $j$ images 
of $\frac{{\sigma}_{\alpha}}{2}$.

{ii) {\it The Case of} $\CP2$}

Coherent states for $\CP2$ can be defined using highest weight
states. For IRR $(3,0)$, we can pick the highest weight state with
$I=I_3=0,\ \ \ Y=-2/3$, namely the ${\lambda}-$quark:
$|0,0,-2/3\rangle$ ${\equiv}$ $|0,0,-2/3;(3,0)\rangle$. Then if
$g{\longrightarrow}U^{(3,0)}(g)$ defines the IRR,
$|g;(3,0)\rangle=U^{(3,0)}(g)|0,0,-2/3;(3,0)\rangle$. For the IRR $(N,0)$, we
can simply replace $|0,0,-2/3;(3,0)\rangle$ by its $N-$fold tensor product
$|0,0,\!-\!2/3;(3,0)\rangle{\otimes}|0,0,\!-\!2/3;(3,0)\rangle
{\otimes}...{\otimes}|0,0,-2/3;(3,0)\rangle$ $=|0,0,-2N/3;(N,0)\rangle$
and set 
\begin{equation}
|g;(N,0)\rangle=U^{(N,0)}(g)|0,0,-\frac{2N}{3};(N,0)\rangle. 
\end{equation}
For $(0,N)$, we
can use the ${\bar{\lambda}}-$quark state 
$|g;(0,3)\rangle$ $=U^{(0,N)}(g)|0,0,+2/3;(0,3)\rangle$
and its tensor product states.

The development of ideas now keep 
following $S^2={\mathbb C}P^1$. Full details
can be found in \cite{ref6}. General theory confirms that the map
$a{\longrightarrow}{\tilde a}$ from matrices in the $(N,0)[(0,N)]$
IRR to functions on ${\mathbb C}P^2$, defined by
${\tilde a}({\xi})=\langle (N,0);g|a|g;(N,0)\rangle({\tilde a}(\xi)
=\langle(0,N);g|a|g;(0,N)\rangle)$
is one-to-one so that a $*-$product on ${\tilde a}$'s exists. In
this map, the $SU(3)$ generators ${\cal L}_i$ acting on
${\tilde a}$ become the corresponding ${\mathbb C}P^2$  $SU(3)$
operators
$-if_{ijk}{\hat\xi}_j\frac{\partial}{{\partial}{\hat\xi}_k}$. 
We shall use the same symbol ${\cal L}_i$ for these
operators too. The orbital $SU(3)$ action is compatible with $*$
in the sense that ${\cal L}_i({\tilde a}*{\tilde b})=({\cal
L}_i{\tilde a})*{\tilde b} + {\tilde a}*({\cal
L}_i{\tilde b})$. Irreducible tensor operators of $SU(3)$
are well studied \cite{SU3tensor}. With their help, fuzzy
analogues of $D-$matrices can be constructed, as also sections
of $U(1)$ and $U(2)$ bundles.

The fuzzy ${\mathbb C}P^2$ Dirac operator is the cut-off version
of (\ref{s511}). It can be put in a matrix 
form as in (\ref{offdiagform1},\ref{offdiagform2}).
We omit the details: the necessary group
theory is already to be found in \cite{ref11} while the
rest is routine.

\vspace{0.2in}
{\bf Acknowledgements}

We have benefited greately from by participating 
in the study groups at CINVESTAV, Mexico D.F. and Syracuse 
during the Summer and Fall of 2000. 
We are especially grateful to Brian Dolan, Xavier Martin,
Denjoe O'Connor and Peter Pre\v{s}najder for their numerous inputs.
This work was supported by DOE and NSF under contract
numbers DE-FG02-85ERR40231 and INT-9908763
repectively and by INFN, Italy.

\appendice

\centerline{Fuzzy $\CP2$ as ``Fuzzy'' Algebraic Variety}

Here we will provide the derivation of Eq. (\ref{fuzzys24})
and derive therefrom the expressions for the 
quadratic Casimir operator $C_2$ in $(N,0)$
and $(0,N)$ representations.

The symmetric representations $(N,0)$ 
of $SU(3)$ that appear in our
$\CP2$ study can be constructed using three creation
operators ${a_i}^\dagger$ and their adjoints $a_i$. We have
\begin{equation}
\left[a_i,{a_j}^\dagger\right]=\delta_{ij},\ \ i,j=1,2,3.
\label{comm1}
\end{equation} 
For the representations $(0,N)$, we need three more creation operators
$b^\dagger_i$ and their adjoints $b_j$. We concentrate below 
on $(N,0)$, the treatment of $(0,N)$ being similar.

The $SU(3)$ generators 
are $\Lambda_a=a^\dagger\,t_a\, a,\ \ t_a=\frac{1}{2}\lambda_a$.
They fulfill 
\begin{equation}
\left[\Lambda_a,\Lambda_b\right]=i\,f^{abc}\,\Lambda_c.
\label{comm2}
\end{equation}  
The Hilbert space ${\cal H}_{(N,0)}$ 
for $(N,0)$ is spanned by states of the form
\begin{eqnarray}
|n_1,n_2,n_3\rangle&=&{a_1^\dagger}^{n_1}{a_2^\dagger}^{n_2}{a_3^\dagger}^{n_3}
|0\rangle,\ \ \ n_1+n_2+n_3=N,\nn\\
a_i|0\rangle&=&0,\ \  i=1,2,3.
\label{states}
\end{eqnarray}
The dimension of this space is
$\frac{1}{2}(N+1)(N+2)$.

Using the definition
$d_{abc} =2\,{\rm Tr}(t^a\{t^b,t^c\})$ the left-hand
side of (\ref{s24}) becomes 
\begin{eqnarray}
d_{abc}\Lambda_b\Lambda_c&=&
2{\rm Tr}(t^a\{t^b,t^c\})\Lambda_b\Lambda_c=\\
&=& 2t^a_{ij}(t^b_{jk}t^c_{ki}+t^c_{jk}t^b_{ki})
\frac{1}{4}a^\dagger_m t^b_{mn} a_{n} a^\dagger_p t^c_{pq} a_{q}
\label{d1}
\end{eqnarray}
The similar expression for the quadratic Casimir $C_2$ is
\begin{equation}
\Lambda_b\Lambda_b=a^\dagger_m t^b_{mn} a_n a^\dagger_k t^b_{kl} a_l
\end{equation}
Taking advantage of the Fierz identity
\begin{eqnarray}
\sum_\alpha (t^\alpha)_{ij}\,(t^\alpha)_{kl}=
\frac{1}{2}\delta_{il}\delta_{jk}- \frac{1}{6}\delta_{ij}\delta_{kl}
\end{eqnarray}
to reduce the summations over the $b$ and $c$ indices,
after a somewhat tedious, but straitforward, computation
one gets
\begin{eqnarray}
\Lambda_b\Lambda_b&=&\frac{1}{2}a^\dagger_m a_n a^\dagger_n a_m
-\frac{1}{6}a^\dagger_m a_m a^\dagger_n a_n
\label{cas2},\\
d_{abc}\Lambda_b\Lambda_c
&=&2t^a_{\alpha\beta}\left[
\frac{1}{4}a^\dagger_l a_\beta a^\dagger_{\alpha} a_l
-\frac{1}{6}a^\dagger_m a_m a^\dagger_{\alpha} a_\beta
-\frac{1}{6}a^\dagger_\alpha a_\beta a^\dagger_l a_l
+\frac{1}{4}a^\dagger_\alpha a_k a^\dagger_{k} a_\beta
\right].
\label{cubic}
\end{eqnarray}
where summation over the
repeated indices is assumed.

At this point we have to use the fact that these operators act 
on the special states that belong to ${\cal H}_{(N,0)}$.
For the states in (\ref{states}), one has 
\begin{equation}
\sum_i a^\dagger_i a_i |n_1,n_2,n_3\rangle=(n_1+n_2+n_3)|n_1,n_2,n_3\rangle=
N|n_1,n_2,n_3\rangle.
\label{constraint}
\end{equation}
From this and (\ref{comm1}) we have the value of the quadratic Casimir
$C_2:C_2=\frac{1}{3}N^2+N$. Using
the fact that $t^a$'s are traceless, we find that the right-hand
side of (\ref{cubic})
when acting on the states from ${\cal H}_{(N,0)}$ becomes
\begin{equation}
d_{abc}\Lambda_a\Lambda_b=2t^a_{\alpha\beta}\left[
\frac{N}{6}+\frac{1}{4}\right]a^\dagger_{\alpha}a_{\beta}=
\left(\frac{N}{3}+\frac{1}{2}\right)\Lambda_a.
\label{cubic2}
\end{equation}

Since ${a^\dagger}_i a_j$ transforms like $(N,0)\otimes(0,N)$,
$\Lambda^L_i$ also fulfills (\ref{cubic2}).\\
With $\hat\xi_i=\Lambda^L_i/\sqrt{N^2/3+N}$, (\ref{fuzzys24}) follows.

There are identical results for the $(0,N)$ representations. The
proofs only involve replacing $a^\dagger_i$ and $a_j$ by
$b^\dagger_i$ and $b_j$ in the preceding discussion.


\appendice

\centerline{Why ${\mathbb C}P^2$ is not Spin}


%

It is a standard result that ${\mathbb C}P^2$ does not admit a
spin structure, but does admit a spin$_c$ structure. 
We plan to explain this result here adapting an argument of 
Hawking and Pope \cite{ref19}. 
The reasoning shows that ${\mathbb C}P^{N}$ for 
any even $N{\geq}2$ is not spin whereas it is spin if $N$ is odd. 

The obstruction to the $\CP2$ spin structure comes from noncontractibile two-spheres in $\CP2$. Since ${\mathbb C}P^2{\simeq}SU(3)/U(2)$, ${\pi}_2({\mathbb C}P^2)={\pi}_1[U(2)]={\mathbb Z}$. Also ${\pi}_1({\mathbb C}P^2)=\{0\}$ so that 
Hurewitz's theorem \cite{ref24} 
leads to $H^{2}({\mathbb C}P^2,{\mathbb Z})={\mathbb Z}$.
Its mod $2$ reduction is $H^{2}({\mathbb C}P^2,{\mathbb Z}_2)={\mathbb Z}_2$.
The absence of spin structure means that the tangent bundle is associated with the non-trivial element of ${\mathbb Z}
_2$. 

Consider a continuous map $g$ of the square $\{(s,t):0{\leq}s;t{\leq}1\}$ 
into $SU(3)$ which obeys the following conditions:
\begin{eqnarray}
&g(s,0)=g(0,t)=g(1,t)=\,{\rm identity}\,{\bf 1},&\\
&g(s,1)=e^{i{\pi}s({\lambda}_3+{\sqrt{3}}{\lambda}_8)}.&
\end{eqnarray}
[See Fig.1.]

\begin{center}
\setlength{\unitlength}{0.5mm}
\begin{picture}(100,80)
\thicklines
\put(-12,12){\makebox(0,0){$(s,t)=(0,0)$}}
\put(-12,85){\makebox(0,0){$(0,1)$}}
\put(112,12){\makebox(0,0){$(1,0)$}}
\put(112,85){\makebox(0,0){$(1,1)$}}
\put(0,20){\line(1,0){100}}
\put(0,20){\line(0,1){60}}
\multiput(0,80)(10,0){10}{\line(1,0){5}}
\put(100,20){\line(0,1){60}}
\put(-5,45){\vector(0,1){15}}
\put(-5,65){\makebox(0,0){t}}
\put(50,15){\vector(1,0){15}}
\put(70,15){\makebox(0,0){s}}
\put(50,5){\makebox(0,0){$g(s,0)=1$}}
\put(-10,50){\makebox(0,0)[r]{$g(0,t)=1$}}
\put(110,50){\makebox(0,0)[l]{$g(1,t)=1$}}
\put(50,95){\makebox(0,0){$g(s,1)=exp(i\pi s(\lambda_3+\sqrt{3}\lambda_8))$}}
\put(50,-10){\makebox(0,0){\bf Fig.1}}
\end{picture}
\end{center}

The curve $g:(s,1){\longrightarrow}g(s,1)$ is a loop in 
$U(2)=\{{\rm stability ~group ~of} ~{\xi}^0\}$ not contractible to identity while 
staying within $U(2)$. It is the generator of ${\pi}_1(U(2))$ and is 
associated with nonabelian $U(2)$ monopoles \cite{extra1}. 
But since ${\pi}_1(SU(3))=\{0\}$, $g$ can be defined 
smoothly in the entire square.

Now $U(2)$ being the stability group of ${\xi}^0$ is contained in the 
tangent space group $SO(4)$ at ${\xi}^0$.
If $x=(x_\mu:\mu=1,2,3,4)$ is a tangent vector 
at $\xi^0$, we can map it to a $2{\times}2$ 
matrix $M(x)=x_4 +i\vec{\tau}\cdot\vec{x}$ 
(${\tau}_i=$ Pauli matrices) with the 
reality property $M(x)^{*}={\tau}_
2M(x){\tau}_2$. $SO(4)=[SU(2){\times}SU(2)]/{\mathbb Z}_2$ acts on $M(x)$ 
according to $M(x){\longrightarrow}h_1M(x)h_2^{\dagger}$, $h_{i}{\in}SU(2)$ 
preserving the reality property and the determinant 
${\rm det}\,M(x)=\sum x_\mu^2$, and hence induces an $SO(4)$ 
transformation on $x$. $U(2)$ is imbedded in this $SO(4)$, 
acting on $M(x)$ as follows: 
$M(x){\longrightarrow}h_1M(x)e^{-i{\tau}_3{\theta}}$.

The spin group $SU(2){\times}SU(2)=\{(h_1,h_2)\}$ is a two-fold cover 
of $SO(4)$. The inverse image of $U(2)$ in $SU(2){\times}SU(2)$ is $SU(2){\times}U(1)$, also a two-fold cover of $U(2)$. In this cover 
the loop $g:(s,1){\longrightarrow}g(s,1)$ becomes $
s{\longrightarrow}(e^{i{\pi}s{\tau}_3},e^{i{\pi}s{\tau}_3})$. 
It is no longer a loop, but runs from $({\mathbb I,\mathbb I})$ to $(-\mathbb I,-\mathbb I)$. 
It is this that obstructs the spin structure, as the following reasoning 
encountered in \cite{ref19} shows.

Let $SU(3){\longrightarrow}\CP2{\simeq}SU(3)/U(2)$ be the map $h{\in}SU(3){\longrightarrow}h{\lambda}_8h^{-1}={\lambda}_{\alpha}{\xi}_{\alpha}$. $U(2)$ here has generators ${\lambda}_i(i=1,2,3)$ and ${\lambda}_8$. This map takes the entire boundary of the 
square $\{g(s,t)\}$
to ${\xi}^0$ and the square itself to a $2-$sphere $S^2$. 

\begin{center}
\setlength{\unitlength}{0.5mm}
\begin{picture}(100,100)
\thicklines
\put(0,20){\line(1,0){100}}
\put(0,20){\line(0,1){60}}
\multiput(0,80)(10,0){10}{\line(1,0){5}}
\put(100,20){\line(0,1){60}}
\put(-10,85){{\bf P}}
\put(110,85){{\bf Q}}
\put(50,10){\makebox(0,0){\bf I}}
\put(-10,50){\makebox(0,0)[r]{\bf II}}
\put(110,50){\makebox(0,0)[l]{\bf III}}
\put(50,90){\makebox(0,0){\bf IV}}
\put(50,-5){\makebox(0,0){\bf Fig.2}}
\end{picture}
\end{center}

Now the tangent space at ${\xi}^0$ of $\CP2$ is spanned by the four $SU(3)$ Lie algebra directions $K^{+},K^{0},\bar{K}^{0},K^{-}$ (in a complex basis). If we write $\CP2$ as $\{h{\lambda}_8h^{-1}\}$, a basis of tangents (a frame) 
at $\xi^0$
is ${\lambda}_a(a=4
,5,6,7)$. Clearly $g(s,t){\lambda_a}g(s,t)^{-1}$ gives a frame 
at $g(s,t){\lambda}_8g(s,t)^{-1}$ of $\CP2$. This gives us a rule for 
transporting this frame (and hence any frame) smoothly 
along curves over $S^2{\in}\CP2$: If $\{(s(\tau),t(\tau)),0{\leq}
\tau{\leq}1\}$ is a curve on the square, the transport of the frame along the curve $ g(s(\tau),t(\tau)){\lambda}_8g(s(\tau),t(\tau))^{-1}$ in $S^2$ is $g(s(\tau),t(\tau)){\lambda}_ag(s(\tau),t(\tau))^{-1}$. In this rule, for the three sides I, II, III (see Fig.2), we 
have $g(s,t){\lambda}_8g(s,t)^{-1}={\lambda}_8$ and $g(s,t){\lambda}_ag(s,t)^{-1}={\lambda}_a$, so that we are at ${\xi}^{0}$ with the frame held fixed. Along side IV, we are still at ${\lambda}_8$ or ${\xi}^0$, but we are rotating $
{\lambda}_a$ 
according to $exp\{{i{\pi}s({\lambda}_3 +{\sqrt{3}}{\lambda}_8)}\}\,{\lambda}_a\,exp\{{-i{\pi}s({\lambda}_3+{\sqrt{3}}{\lambda}_8)}\}$, it is a $2{\pi}-$ rotation of the frame as $s$ varies from $0$ to $1$.

If spinors can be defined on $\CP2$, this transport of frames will 
consistently lead to their transport as well. Thus along sides I, II, 
III, we should be able to pick a suitable constant spinor ${\psi}$. 
But then, along IV, as $s$ increases to $1$, we will arrive at $Q$ with 
$-{\psi}$ 
as $(-{\bf 1},-{\bf 1})$ of $SU(2){\times}SU(2)$ flips the sign of a spinor. 
As we had ${\psi}$ along III, we lose continuity at $P$ and find that 
spinors do not exist for $\CP2$.

It is possible to show that this conclusion is not sensitive to our 
choice of rule of transport of 
frames (that is, connection in the frame bundle).

The ${\rm spin}_c$ structure is achieved by introducing an additional $U(1)$ 
connection for spinors which amounts to adding a hypercharge of 
magnitude $1$. That would give an additional 
phase $exp(i{\pi}{\sqrt{3}}{\lambda}_8s)$ along IV and an extra minus 
sign at $s=1$ cancelling the above unwanted 
minus sign. Note that $1)$ this connection and extra hypercharge 
cancels out for frames which contain a spinor and a complex conjugate spinor, 
$2)$ there is no vector bundle with this extra connection as 
its existence gives a
contradiction just as does the existence of the spin bundle.

Let us see what all this means for $SU(3)$. Under $U(2)$, at ${\xi}^0$, 
the tangents transform as $K's$ and $\bar{K}'s$, that is as the IRR's $(I,Y)=(\frac{1}{2},1)$ and $(\frac{1}{2},-1)$. From the way $M(x)$ transforms, we can see that $Y$ corresponds to ${\tau}_3$ where ${\tau}_{\alpha}/2$ are 
$SU(2)$ generators acting on the right of $M(x)$.

The $SU(2){\times}SU(2)$ IRR's of the non-existent spinors are as follows: 
$i)$Left-handed spinors: $(1/2,0)$, $ii)$right-handed spinors:
$(0,1/2)$.
The corresponding $(I,Y)$ quantum numbers are thus: 
$i)$ left-handed spinors: $(1/2,0)$, $ii)$ right-handed spinors: $(0,1)$ 
and $(0,-1)$. 
The quantum numbers in the ${\rm spin}_c$  case follows by adding an 
additional hypercharge which we can take to be $-1$: 
$1)$ Left-handed 
${\rm spin}_c$ : $(1/2,-1)$. $2)$ Right-handed ${\rm spin}_c$ : $(0,0)$ and 
$(0,-2)$. These are precisely the $U(2)$ quantum numbers of the representation 
space of tangent ${\gamma}'s$ in Section $6$. The $SU(3)$ 
IRR's have to contain these $U(2)$ IRR's. They are not symmetric between 
left- and right-handed spinors.

The ${\rm spin}_c$  structures are not unique. Thus we have the freedom to add 
additional hypercharge $2n$ $(n{\in}Z)$ to the ${\rm spin}_c$  spinors, 
that is, 
tensor the ${\rm spin}_c$  bundle with any $U(1)$ bundle. 
The choice of ${\rm spin}_c$  
in our text is natural for our Dirac operator.

{\it On General ${\mathbb C}P^N$}

${\mathbb C}P^N$ for all odd $N$ admits a spin structure whereas those for even $N$ admit only a ${\rm spin}_c$  structure \cite{ref25}. 
We can understand this result too by pursuing the preceding arguments.



Let $Y^{(N+1)}=\frac{1}{N+1}\,{\rm diag}\,(1,1,...1,-N)$ be the $SU(N+1)$
``hypercharge''. The previous $Y$ is $Y^{(3)}$. We can represent ${\mathbb C}P^{N}=SU(N+1)/U(N)$ as $\{hY^{(N+1)}h^{-1}:h{\in}SU(N+1)\}$, the stability group $\{u{\in}SU(N+1):uY^{(N+1)}u^{-1}=Y^{(N+1)}\}$ being $U(N)$.

For all $N{\geq}1$, the square of Figs.1 and 2 and the map 
$g:(s,t){\longrightarrow}g(s,t){\in}SU(N+1)$ can be constructed so 
that it is constant on sides I,II and III while
$g:(s,1){\longrightarrow}g(s,1)$ gives a generator of ${\pi}_1 (U(N))$. There is obstruction to spin structure if this loop 
when it acts on a frame at $Y^{(N+1)}$ rotates it by $2{\pi}$, that is acts as the noncontractible loop of $SO(2N)$.

Let $(q_1,q_2,...,q_{N+1})$ be the ``quarks'' of $SU(N+1)$. 
The hypercharge $Y^{(N)}$ of $SU(N)$ acts as the 
generator $\bar{Y}^{(N)}=\frac{1}{N}(1,1,..,-(N-1),0)$ on these quarks. 
We can choose the loop $g:(s,1){\longrightarrow}g(s,1)$ according to
\begin{eqnarray}
g(s,1)&=&e^{i\frac{2{\pi}s}{N}(N\bar{Y}^{(N)})}e^{-i\frac{2{\pi}s}{N}(N+1)Y^{(N+1)}}\nonumber\\ 
&=& \left[\begin{array}{ccccc}
                          1&0&.&.&0\\
                          0&1&0&.&0\\
                           0&.&.&1&0\\
                           0&.&.&e^{-i2{\pi}s}&0\\
                           0&0&0&0&e^{i2{\pi}s}
           \end{array}
     \right].
\end{eqnarray}
The tangent vectors at $Y^{(N+1)}$ transform 
like $\bar{q}^{(i)}q^{(N+1)}$ and $\bar{q}^{(N+1)}q^{(i)}(1{\leq}i{\leq}N)$. 
So under $g(s,1)$,
\begin{eqnarray}
\bar{q}^{(i)}q^{(N+1)}&{\longrightarrow}&e^{i2{\pi}s}\bar{q}^{(i)}
q^{(N+1)},i\leq N-1,\nonumber\\
\bar{q}^{(i)}q^{(N+1)}&{\longrightarrow}&e^{i4{\pi}s}\bar{q}^{(i)}
q^{(N+1)}\ \ {\rm for}\ \ i=N,\nonumber\\
\bar{q}^{(N+1)}q^{(i)}&{\longrightarrow}&e^{-i2{\pi}s}\bar{q}^{(N+1)}
q^{(i)},i{\leq}N-1,\nonumber\\
\bar{q}^{(N+1)}q^{(i)}&{\longrightarrow}&e^{-i4{\pi}s}\bar{q}^{(N+1)}
q^{(i)}\ \ {\rm for}\ \ i=N.
\end{eqnarray}
Each $i$ gives a plane in $2N$ dimensions and each 
factor $e^{i2{\pi}s}$ in the first two lines
gives a $2{\pi}-$rotation.

Thus we have a product of $(N-1)+2=(N+1)$ $2{\pi}-$rotations. For $N$ odd, they are contractible in $SO(2N)$, and for $N$ even, they are not, showing the result we were after.

\bibliographystyle{unsrt}

\begin{thebibliography}{99}
\bibitem{ref20}
The analysis of the $SU(3)$ representation content
of $V(g)$ is due to Brian Dolan.
\end{thebibliography}

\end{document}